\documentclass[10pt,final,journal,letterpaper,twoside]{IEEEtran} %% For final form
\usepackage{float}
\usepackage{algorithm}
\usepackage{algorithmic}
\usepackage{subcaption}
\usepackage{graphicx}
\usepackage{gensymb}
\usepackage{balance}
\usepackage{amsmath}
\usepackage{comment}
\usepackage{tabularx}
\usepackage{ifthen}
\usepackage{cite}
\usepackage{subcaption} 
\usepackage{multirow} % For multirows
\usepackage{amssymb}
\usepackage{cleveref}
\usepackage{array}
\usepackage{balance}
\begin{document}
%----------------------------------------------------------------------
% Title Information, Abstract and Keywords
%----------------------------------------------------------------------
\title{IMU-based Modularized Wearable Device for Human Motion Classification}
\author{Sahan~Wijethunga,
    Shehan~Kaushalya~Senavirathna,
    Kavishka~Dissanayake,
    Janith~Bandara~Senanayake,
    Eranda~Somathilake,
    Upekha~Hansanie~Delay,
    Roshan~Indika~Godaliyadda,~\IEEEmembership{Senior ~Member,~IEEE},
    Mervyn~Parakrama~Ekanayake,~\IEEEmembership{Senior~Member,~IEEE},
    Janaka~Wijayakulasooriya,~\IEEEmembership{Senior~Member,~IEEE}
    }
%\author[Short Names]{First
%Author\member{Student
%Member}\authorinfo{Department
%of Electrical Engineering\\
%Some University, Somewhere CA,
%90210, USA}%
%\and{}Second Author\member{Senior
%Member}\authorinfo{Department
%of Electrical Eng...}
%\and{}and Third
%Author\member{Fellow}\authorinfo{...}
%}
%\journal{IEEE Transactions on Instrumentation and Measurement}

\maketitle

\begin{abstract}
Human motion analysis is used in many different fields and applications. Currently, existing systems either focus on one single limb or one single class of movements. Many proposed systems are designed to be used in an indoor controlled environment and must possess good technical know-how to operate. To improve mobility, a less restrictive, modularized, and simple Inertial Measurement units based system is proposed that can be worn separately and combined. This allows the user to measure singular limb movements separately and also monitor whole body movements over a prolonged period at any given time while not restricted to a controlled environment. For proper analysis, data is conditioned and pre-processed through possible five stages namely power-based, clustering index-based, Kalman filtering, distance-measure-based, and PCA-based dimension reduction. Different combinations of the above stages are analyzed using machine learning algorithms for selected case studies namely hand gesture recognition and environment and shoe parameter-based walking pattern analysis to validate the performance capability of the proposed wearable device and multi-stage algorithms. The results of the case studies show that distance-measure-based and PCA-based dimension reduction will significantly improve human motion identification accuracy. This is further improved with the introduction of the Kalman filter.  An LSTM neural network is proposed as an alternate classifier and the results indicate that it is a robust classifier for human motion recognition. As the results indicate, the proposed wearable device architecture and multi-stage algorithms are cable of distinguishing between subtle human limb movements making it a viable tool for human motion analysis.

\end{abstract}


%----------------------------------------------------------------------
% SECTION I: Introduction
%----------------------------------------------------------------------
\section{Introduction}

\IEEEPARstart{}{Human} Motion Analysis (HMA) has been a dominant field of research in recent history\cite{HMA}. It plays a major role in several applications such as a diagnostic tool in medicine\cite{HMAmedi}, athletic performance analysis in sports \cite{hma_sports}, man-machine interface in different industries \cite{hma_human_machine}, video surveillance \cite{hma_surveillance}, physical therapy\cite{hma_physio}, and kinesiology\cite{inproceedings}. Human motion analysis (HMA) has several sub-sections such as activity and gesture recognition, population motion tracking, individual limb motion analysis, gait analysis, etc. 

Vision-based systems are the most popular among human motion capturing methods as they have the capability to monitor activities of large scale such as limb motion to activities of minuscule scale such as pupil dilation depending on the application requirement. Additionally, cameras will be used as the sensor and they will impose minimal interference to the activity being monitored. However, these vision-based systems are not an ideal solution for HMA, specifically when prolonged monitoring in natural scenes is required. They have several limitations such as the cost of implementation, the algorithmic burden of image processing, and environmental dependency nature. Furthermore, continuous coverage of the working area of the person being tracked is required with no blind spots and this is a specific concern when it comes to highly mobile workers who work in the crowded, large working area. Dedicated focus which enables capturing essential details might be missed when using an all-purpose camera-based system for human motion tracking. As a method of overcoming these challenges faces by vision-based systems, more and more wearable type HMA are being introduced.

Recently a significant amount of research has been performed into the application of wearable sensor-based systems for the measurement of the quantity and quality of the physical activities performed. Inertial sensors such as Micro Electro Mechanical System (MEMS) accelerometers and rate gyroscopes are now widely used for this application. In a study conducted \cite{IMUVsOptical} the performance of Inertial Measurement Sensors (IMUs) and optical motion capture systems were compared. It was concluded that IMU sensors combined with a well-designed algorithm are more reliable in measuring human motion data in a natural environment with higher sensitivity and accuracy. Another study \cite{IMUSinVsMul} have concluded that utilizing multi-sensors for HMA will allow the user to adopt lightweight algorithms and that they are more appropriate in recognizing dynamic activities as well as transition activities. Furthermore, studies were conducted on implementing IMU sensors for different types of applications such as military\cite{military}, sports\cite{Sports} as well as safety\cite{safety}.

Predicting and monitoring HMA via IMUs for the purpose of potential work-related musculoskeletal disorders (MSD) is a less explored area in the current literature. As mentioned in \cite{MSDcashew} MSD can have severe detrimental effects on several components of the human body which include muscles, joints, ligaments, tendons, nerves, and even the blood circulation system. According to several studies conducted all over the world \cite{MSDcashew,MSDchina,MSDiran,MSDiranSugar,MSDsrilanka} MSD are quite prominent among factory workers as well as in other professions such as nursing\cite{MSDnurse}. In a study conducted in china \cite{MSDchina} it was observed that female factory workers as well as workers with more than 55 work hours per week face higher risk related to MSD. Additionally, depending on the type of work, the affected areas of the body also vary. According to \cite{MSDiran} the main ergonomic issues which resulted in MSD are awkward postures, material handling as well as long hours. Further, according to \cite{MSDvsSafetyE} the design and the implementation of safety equipment may also have an effect on workers' risk of MSD. 

The proposed dedicated, less obstructive, low-cost IMU-based wearable device which can monitor and track human motion activities for a prolonged duration in a large work area is ideally suited for this type of HMA-based study that focused on predicting potential work-related MSDs. 
The main goal of this paper is to present the IMU sensor-based device and its capability as a potential wearable for the aforementioned use of possible detection of MSD hence the primary focus of the paper is through case studies to show how subtle hand movements as well as changes in walking patterns due to environmental and shoe type accessory conditions can be detected through the proposed wearable and its affiliated algorithms. 

Thus, in the first half of the paper, the process of designing, conditioning, and fabrication of the system is presented and in the second half its representation, and performance was analyzed utilizing different types of human motions. Therefore, the two case studies considered were selected to highlight two different capabilities of the wearable device and the proposed algorithm for HMA. The first activity is hand gestures. During this, a single module of the device was used with the central control unit to record several different hand gestures and classify them. The second activity was observing the effect of environmental changes on gait while walking. In this, both shoe parameters and the walking environments were changed while utilizing four modules to monitor arms and legs. These case studies have permitted us to observe the effect of human movements that are focused on one particular part of the body as well as ones that affect the entire body. 

With these activities, it can be demonstrated that the device presented in this paper has the ability to monitor whole body movements as well as slight movements of a single limb such as wrist and finger movements. Additionally, in the first case study, the data will be used to classify the types of motion with subtle changes while the second case study's goal is to monitor the changes in the same activity due to external factors such as changes to the environment and shoe type accessories.

Most of the HMA devices that are available, are technically rich and require a higher technical knowledge to operate and obtain measurements. The device we propose will have a simple design and simple instructions to operate and obtain measurements, such that a person with less technical know-how will be able to operate it easily for instance where studies on MSDs potentially be required. The contributions of this study are summarized as follows:
\begin{enumerate}
    \item Design and development of a measurement device that can acquire real-time MEMS sensor reading throughout the body with optimized and robust data acquisition algorithms.

    \item A methodology for the selection of multi-sensory data to be optimized for given classification problems in terms of clustering quality.

    \item Use of the Kalman filter as an active smoother and dimension reduction technique tool. 

    \item A methodology for sensor screening based on measurement power.

    \item Introduction of a distance measure-based dimension reduction technique for high-level reduction of high-dimension temporal data.

\end{enumerate}

The aforementioned contributions allow the wearable device to be used for a long-duration HMA in a dedicated and non-intrusive fashion in work environments for workers with low technical know-how. 
In addition, the contributions have been used for the two case studies, hand gesture recognition and walking pattern analysis in different environmental conditions and shoe parameters using a wide array of Machine Learning (ML) and Deep Learning (DL) algorithms to demonstrate their capability.  It can be noted that the devices have the capability to monitor and capture the effects of repetitive tasks carried out for a long time period.

%----------------------------------------------------------------------
% Background
%----------------------------------------------------------------------

\section{Materials and Methods}\label{sec:M&M}

\subsection{Device}\label{sec:device}
The design and development of the device were done in 3 phases. They are; physical design, circuit design, and coding. The specifics of each individual phase are discussed in this section.

\subsubsection{Phase I : Physical Design}\label{sec:phy}

This device will be mainly implemented in a work-related environment. Our goal was to come up with a physical design versatile enough to handle different types of working environments such as indoor factories, indoor offices, outdoor construction sites, and outdoor spaces. Further, it was designed such that it allows the user to perform various types of activities, such as hand motion, upper body movements,  walking, climbing, jogging, 

The device needs to be wearable, flexible, and agile in order for it to be used in different working environments. Additionally, the aim of this device is to be used in different arrays of applications. Thus rather than designing the system to work in unison, it was designed with a modularized architecture while leaving the capacity to expand and include more sensors and modules if required. 

The main design consists of a single central unit and several modules focusing on a different section of the body. Each module will have four or six sensing units and they could be of any variety, with the ability to easy removal or change sensors such as IMUs, pressure sensors, force-sensitive resistors(FSR), etc. As concluded in the article \cite{IMUbest}, IMUs are one of the best low cost least invasive methods for HMA. Thus the initial modules which are presented in this paper only consist of IMU sensing units. As the primary focus was on testing the feasibility of the device for limb movement analysis with goals such as the identification of MSD as a potential application, the initial case studies focused on limb movements. Therefore, initially, four modules were used to monitor limb motion, with two identical modules for arms and two for legs. The MPU 6050 IMU sensor was used with four sensors per arm module and six sensors per leg module. However, if required additional modules for areas such as the head, neck, and spine can also be included. 

\begin{figure}
    \centering
    \includegraphics[width=0.25\textwidth]{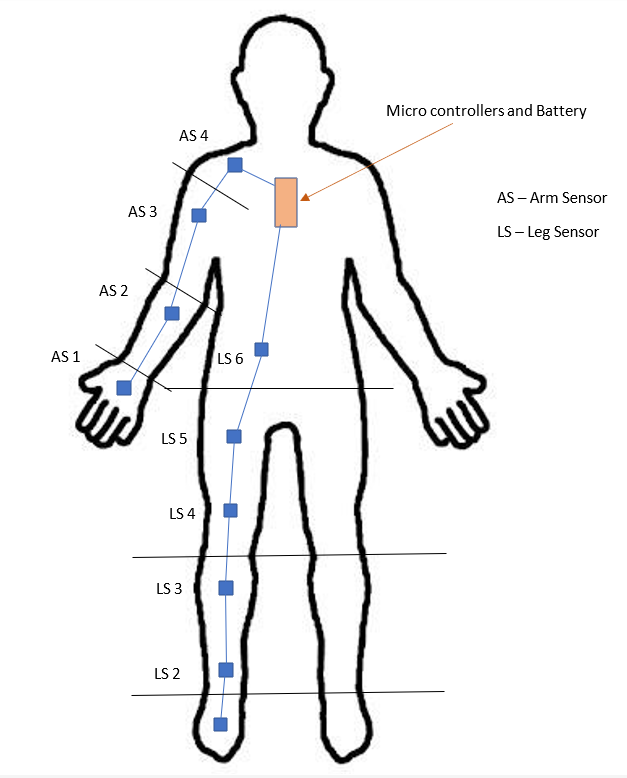}
\caption{Sensor placement on human body}
\label{fig:placing}
\end{figure}

\begin{figure}[!t]
\begin{center}
\includegraphics[width=0.48\textwidth]{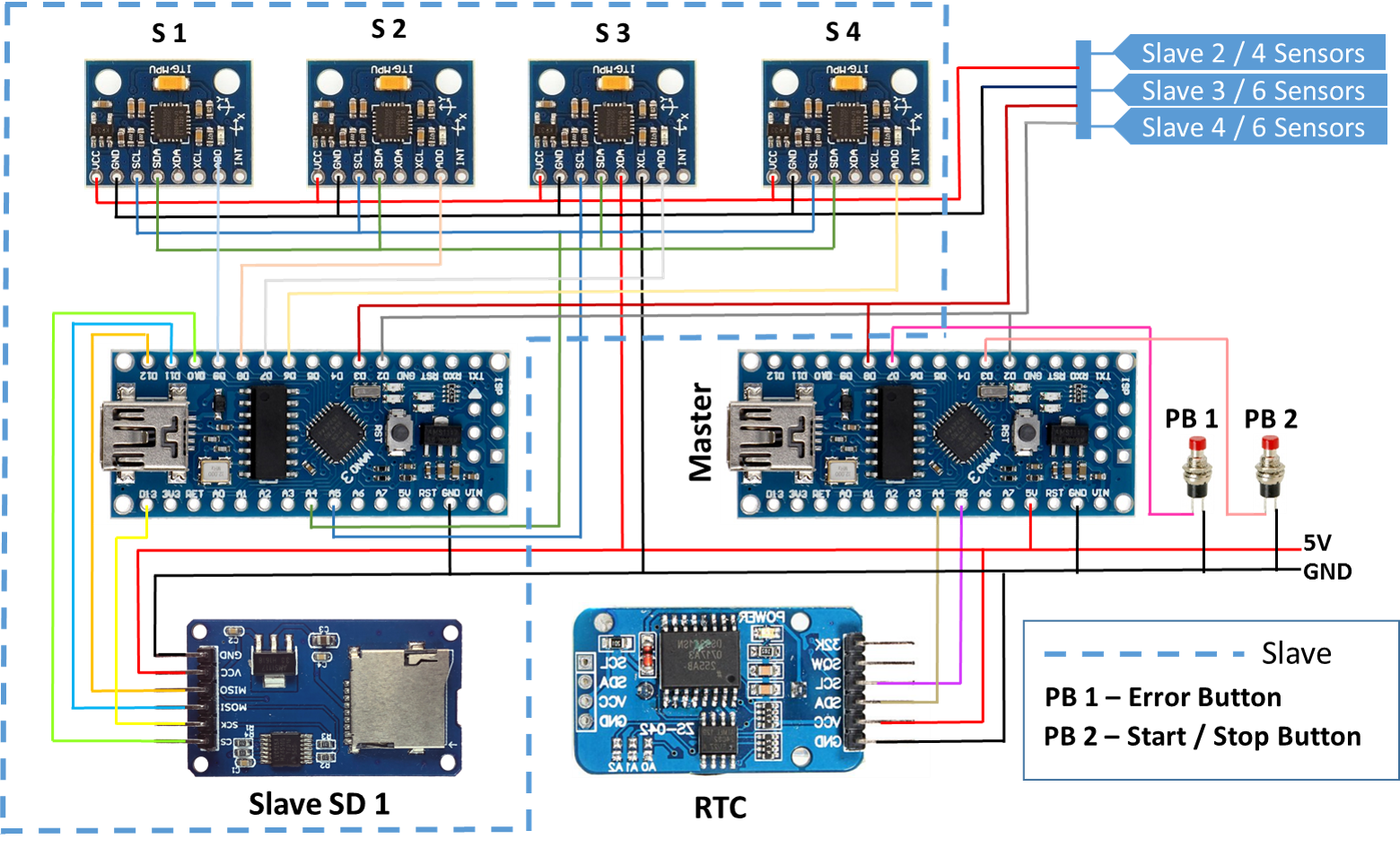}
\caption{Circuit architecture}
\label{fig:circuit}   
\end{center}
\end{figure}
\textbf{Central Unit: } The main purpose of the central unit is to monitor and synchronize the data flow of the sensor. Additionally, it acts as the main power source. This unit comprises an Arduino-nano board and a Real Time Clock (RTC) coupled with a power bank unit. Multiple modules can be connected to this unit. The central unit will synchronize the readings taken from sensors and will send necessary trigger signals for the modules to take reading as well as write the collected data into the SD card.

\textbf{Modules: }Each module may comprise an Arduino-nano controlling board coupled with an SD card unit and have the ability to collect readings from six or less number of sensing units. To maintain the independent nature of each module as well as to reduce the computational burden on the central unit it was decided to include a microcontroller-based unit in each module. Considering the physical size, compatibility with the central unit as well as performance, the Arduino-nano board was selected and an SD card unit was incorporated into each module. The data pertaining to each module was to be stored within itself resulting in less complexity when handling data from multiple sensor systems.

The sensor placement of each module was done such that maximum data of the entire limb can be captured while using the minimum number of sensing units \cite{IEEEtrans1}. The sensor placement of the arm unit and the leg unit can be observed in Figure \ref{fig:placing}. Size adjustable and high elastic fabrics with Velcro straps were used to secure the sensors in place. This method is used to permit different people with different body measurements to use the same setup. Additionally, this allows the unit to be worn and taken off multiple times with ease without damaging the securing setup. The elasticity of the straps accommodates the contraction and relaxation of muscles without moving the sensor position along the limb.

\begin{figure}[!t]
     \centering
     \begin{subfigure}[b]{0.18\textwidth}
         \centering
         \includegraphics[width=\textwidth]{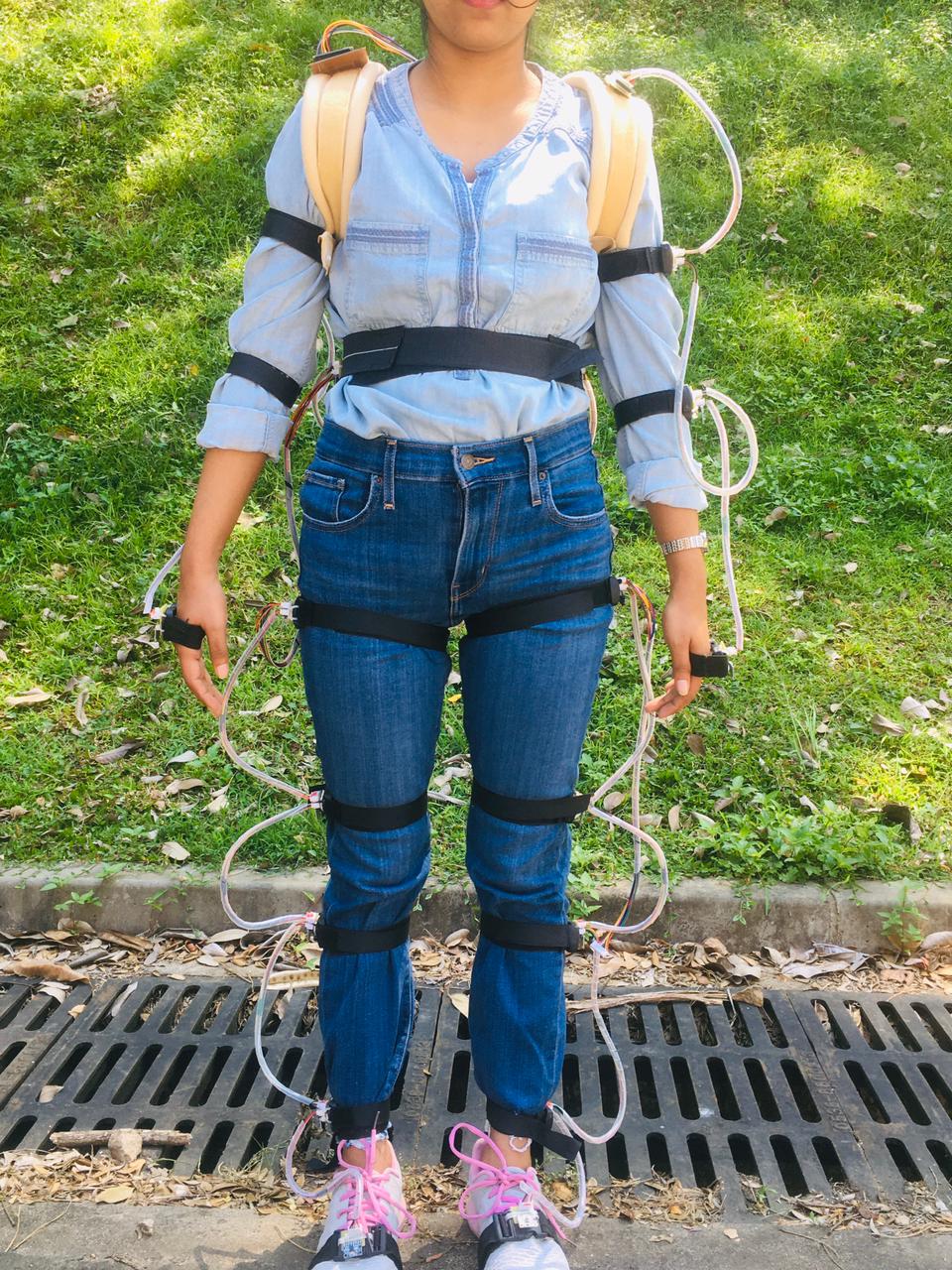}
         \caption{Front View}
         \label{fig:front_view}
     \end{subfigure}
     \begin{subfigure}[b]{0.18\textwidth}
         \centering
         \includegraphics[width=\textwidth]{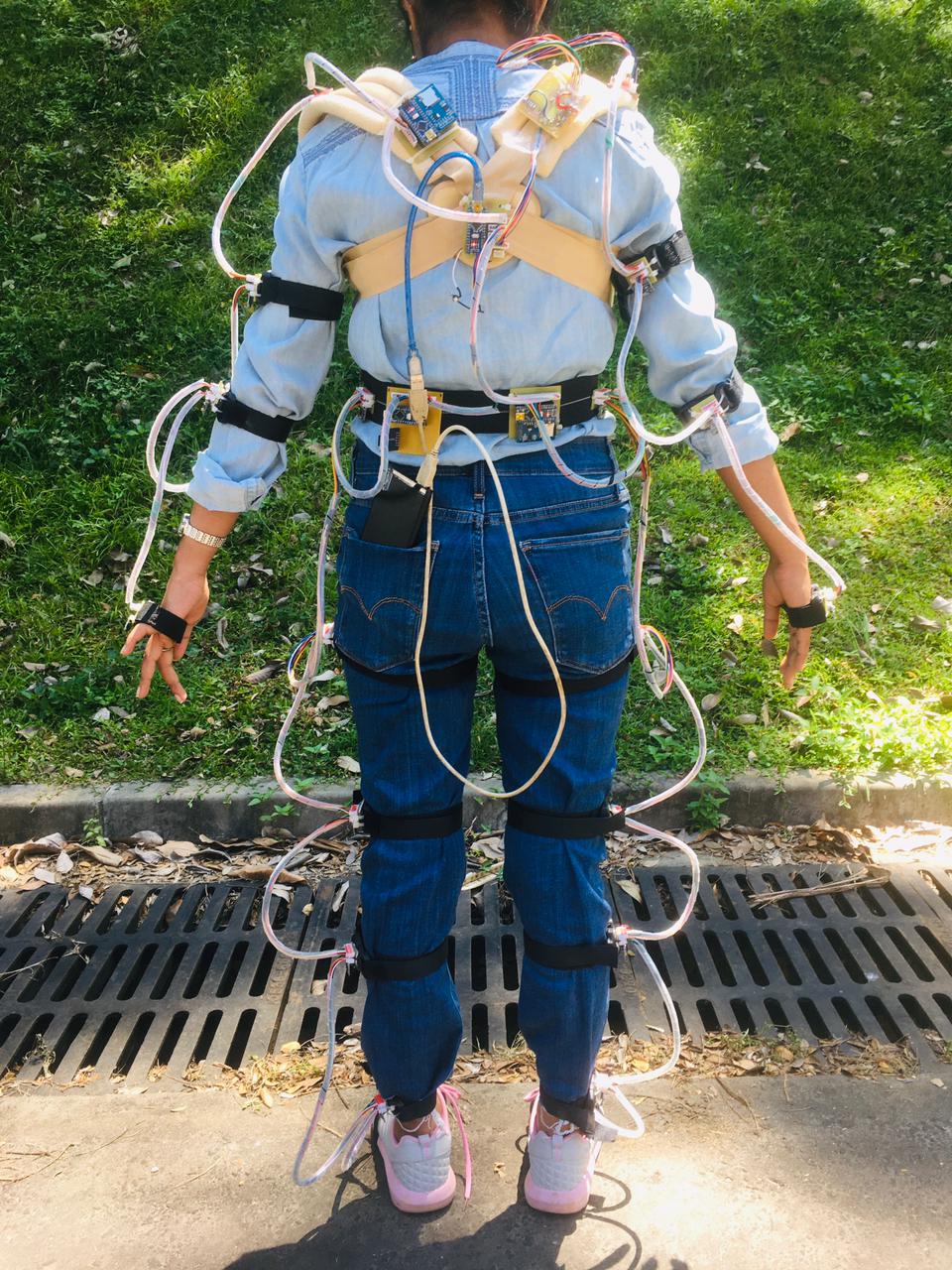}
         \caption{Back View}
         \label{fig:back_view}
     \end{subfigure}
      \begin{subfigure}[b]{0.22\textwidth}
         \centering
         \includegraphics[width=\textwidth]{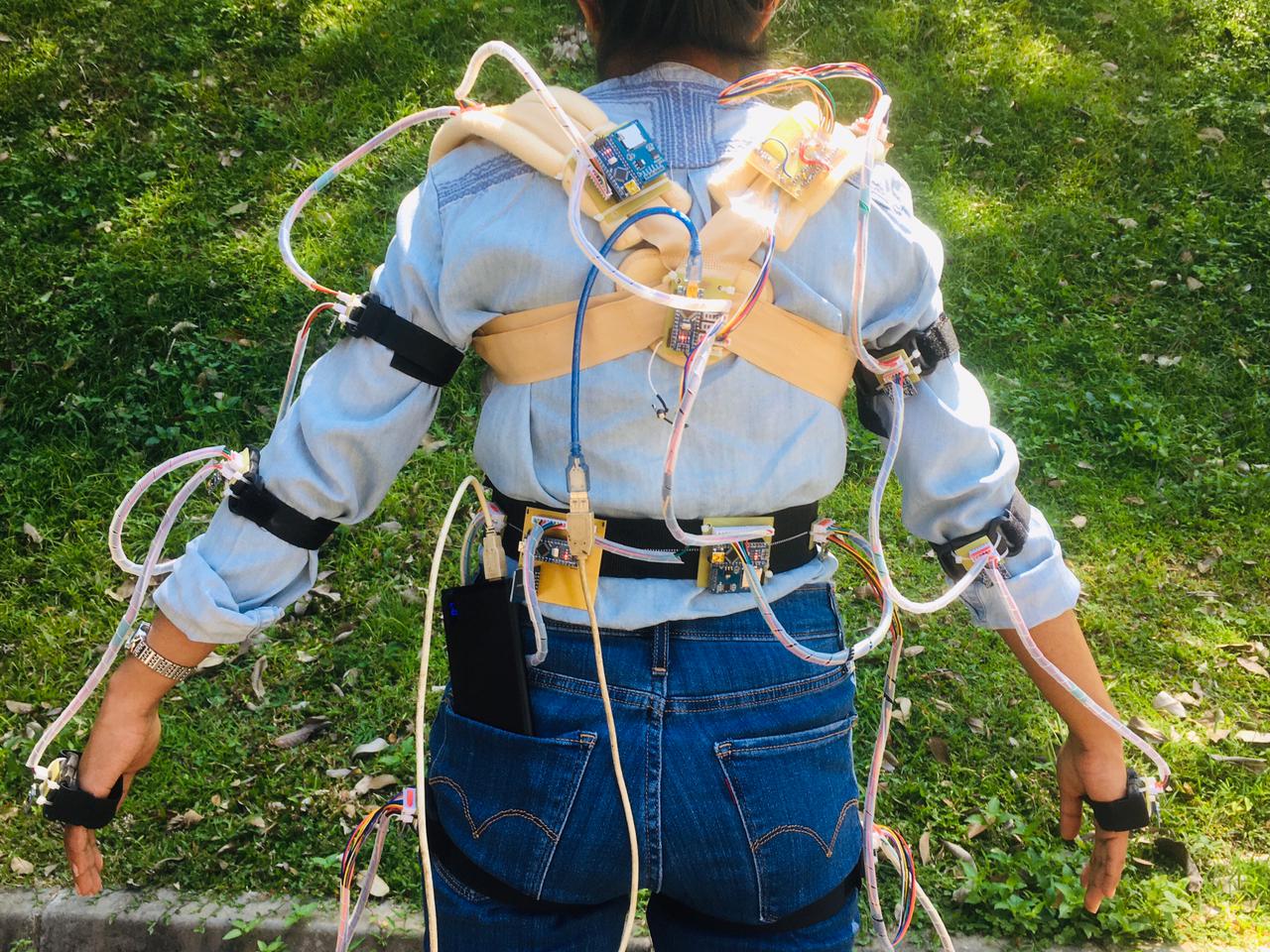}
         \caption{Sensors on Hand}
         \label{fig:hand}
     \end{subfigure}
     \begin{subfigure}[b]{0.15\textwidth}
         \centering
         \includegraphics[width=\textwidth]{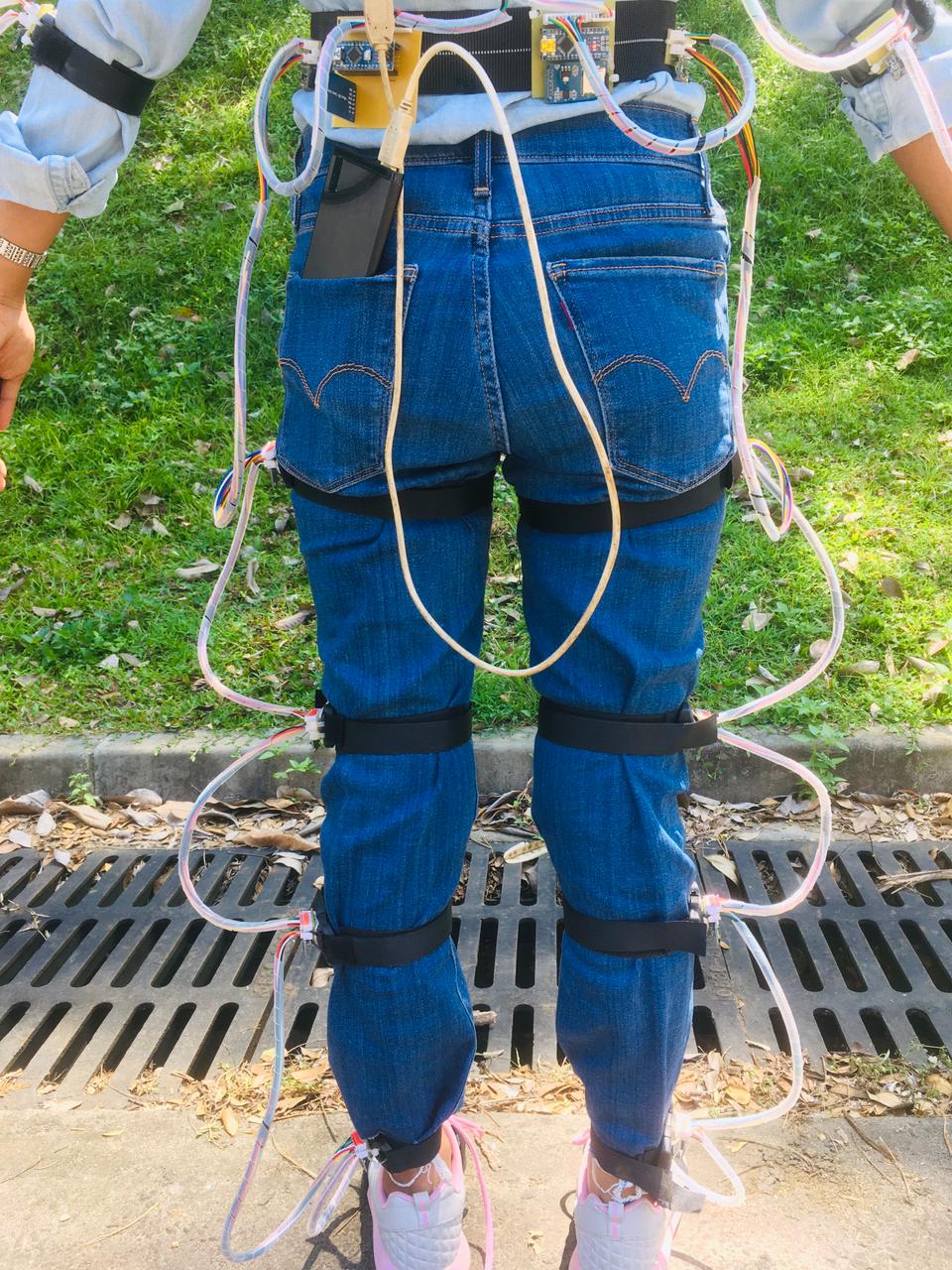}
         \caption{Sensors on Legs}
         \label{fig:legs}
     \end{subfigure}
    \caption{Wearable Device}
    \label{fig:three graphs}
\end{figure}

\subsubsection{Phase II : Circuit Design}\label{sec:cir}

The circuit of this setup was quite simple and efficient. The main circuit diagram of the setup can be observed in Figure \ref{fig:circuit}. The main components of the circuit are discussed in phase 1. A 5V power supply was given to all the microcontrollers and to the SD card units and IMU sensors. Two push buttons are included in the setup. The first one is set up to be the start/stop button. A single press of this button will prompt the central unit to start sending out the trigger signals for the modules, which will prompt each module to start taking readings. And the next press will terminate this process. The second push button is to mark any important occurrences during the data collection period. Additionally, the LED light of the Arduino-nano board was used as an indicator of the status of the SD card.

\subsubsection{Phase III: Coding}

Several considerations and allocations were made during the coding process for both central and module microcontrollers. They will be discussed in this section.

When using multiple sensors usually readings are taken in a loop format \cite{loop1, loop2}. However, in cases such as this where twenty sensors were used if a loop format is implemented the data capturing frequency may become too low. However, this issue is not present in our device due to its modularized architecture. The readings are captured by each module separately and the maximum number of sensors within a module is six. This allows for a higher data acquisition frequency in comparison to non-modularized devices.

Since the modules were designed to operate independently, it was crucial to synchronize the readings taken from each module. This was done using the central unit. Initially, several issues were identified, and subsequently, possible solutions were implemented. Following major issues were faced during the design process of the device. 

\begin{enumerate}
    \item Variation of data capture time due to the variation in data size resulting in shifts in data flow and asynchronous modules, hindering the ability to data merging and overall body motion analysis.
    \item Variation in data capture starting time may include shifts in data
    \item Impossibility of predicting the exact time taken for read-write data cycle of an SD card. 
\end{enumerate} 

\begin{figure}[!t]
\centering
\includegraphics[width=0.4\textwidth]{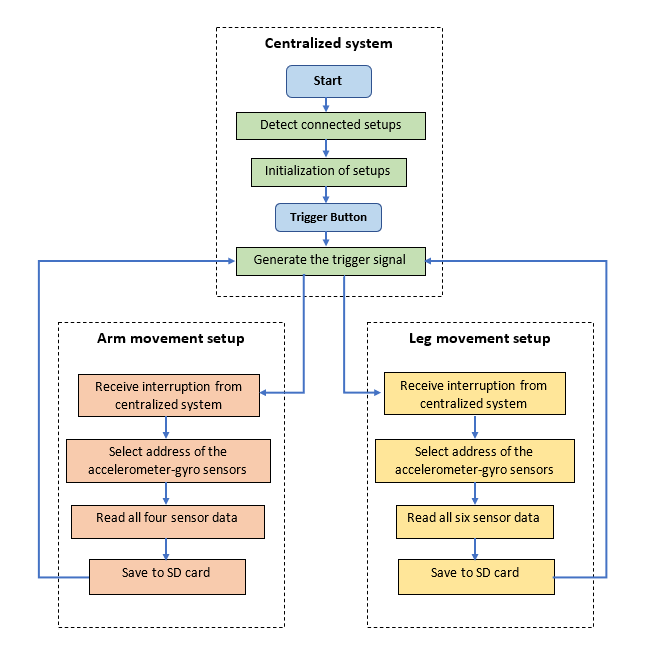}
\caption{Timing diagram}
\label{fig:timing}
\end{figure}

The following solutions were implemented to solve problems one and two. The central unit is used as a triggering device for the modules to synchronize. For this initiative, in the central Arduino, rather than using loops or time-read commands, the inbuilt timer is used to maintain uniform sampling with precise timing. The activities of a single cycle can be observed in Figure \ref{fig:timing}. At the beginning of each cycle, a trigger signal generated by the central unit is sent to the modules. This will prompt all the modules to capture the reading of the sensor one at the same time. Then a counter is started and when it hits the value 4800, the next trigger signal is sent out from the central unit to all the modules to capture the data of sensor two simultaneously. With this, the counter value will revert back to zero and will start to count once more. Again when the counter value hit 4800, another trigger signal will be sent out to capture data from sensor three. This will repeat until triggers are sent to capture data from all six sensors. As observed in Figure \ref{fig:timing} the final trigger sent to capture data from sensor six also contains a trigger to write the collected data. Depending on the size of the data captured the modules may or may not write the data. Since the writing time of the SD card unit is higher, more time is allocated for this action. There is approximately a 2.4ms time gap between triggers and approximately 19.25ms was allocated to write the data

Sending trigger signals to the modules to take readings from sensors simultaneously solved the issue of the shift due to the difference in data size as well as the issue of starting time difference. And allocating a separate time slot to write the data resolved the issue of time shifts due to nonuniform writing frequency. Implementing the above method while coding ensures that all our modules were synchronized and performed in a uniform manner. Additionally to start and end the data capturing a different trigger signal which was prompted by the push button was used.

In the Arduino nano of each module, the sampling frequency is set to 32Hz. The inability to know the status of the SD card was problematic for proper measurements. As a solution, a small program was written to indicate the status of the SD card using the LED light of the Arduino board. Following are the indicated statuses and their signature.

\begin{enumerate}
  \item \textbf{Ready:} Two blinks
  \item \textbf{Start/Stop:} One blink
  \item \textbf{Data capturing in progress:} Continuous fast blink
  \item \textbf{Error: }Continuous slow blinking
\end{enumerate}

In each MPU 6050 sensor, the temperature sensors were disabled permanently to save power as the readings from them are not useful for our purpose at the moment. Additionally, when data capturing is not taking place the sensors are put into sleep mode. Since there is considerable overhead to convert the raw reading to standard units, the MPU6050 library was not used. Instead, raw register values were read using the SPI library and the burst mode was used resulting in unaltered register values being saved in the SD modules.

\begin{figure}[!t]
\centering

\begin{subfigure}[t]{0.138\textwidth}
    \includegraphics[width=1\textwidth]{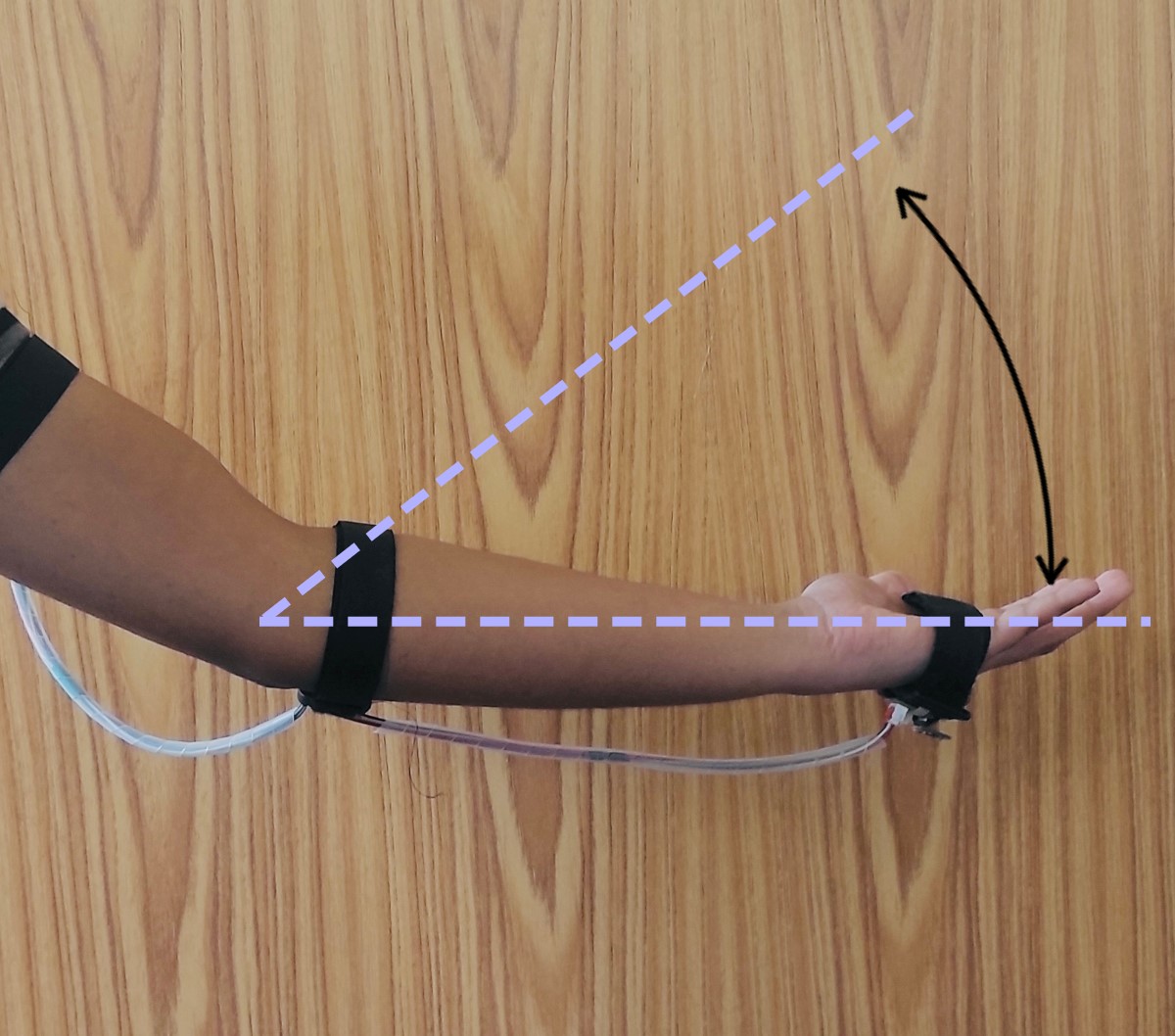}
        \caption{Elbow Flexion and Extension (H1)}
        \label{fig:A}
\end{subfigure}
\hspace{0.5cm}
\begin{subfigure}[t]{0.12\textwidth}
    \includegraphics[width=1\textwidth]{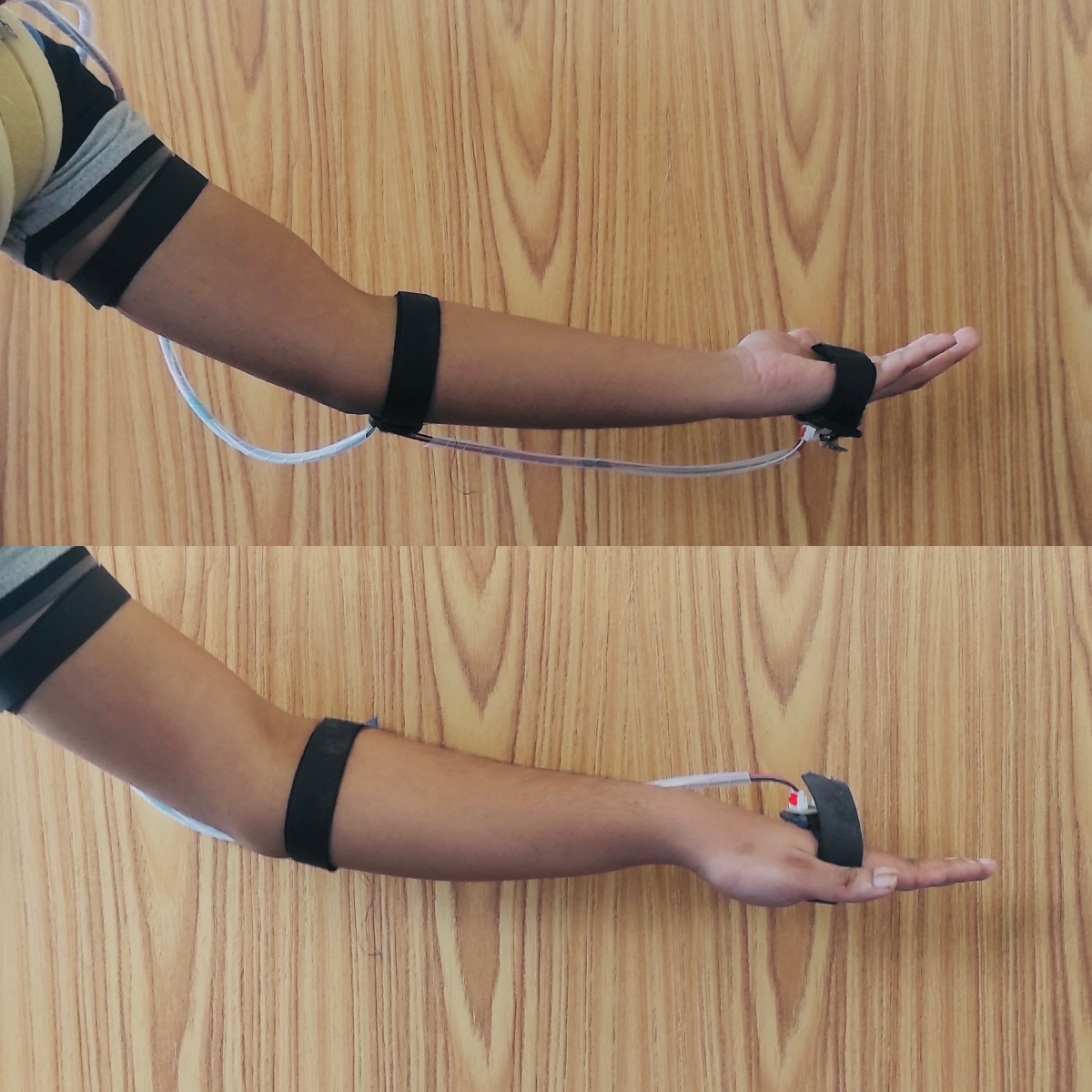} 
        \caption{Elbow Pronation and Supination (H2)}
        \label{fig:B}
\end{subfigure}
\hspace{0.5cm}
\begin{subfigure}[t]{0.12\textwidth}
    \includegraphics[width=1\textwidth]{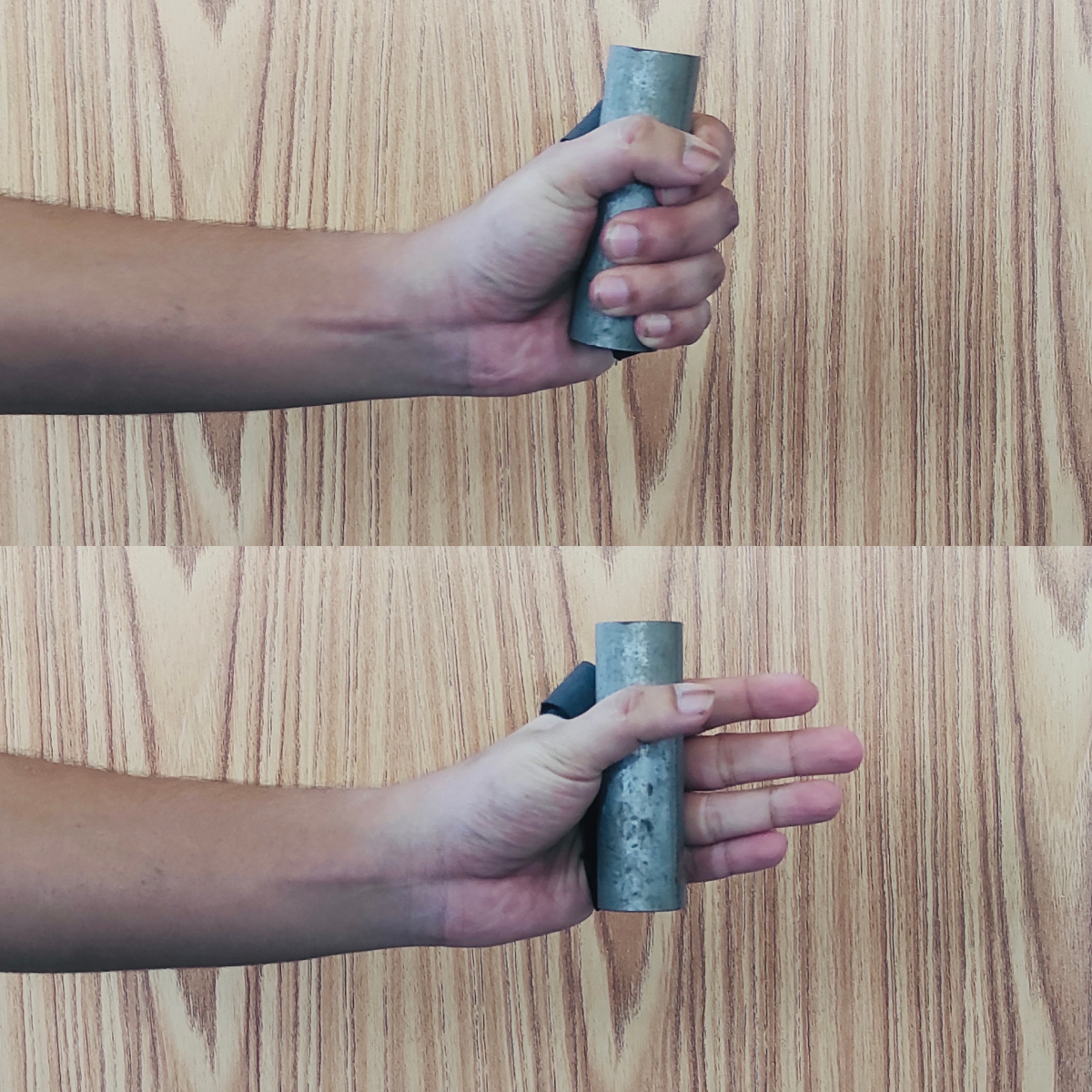} 
        \caption{Power Grip (H3)}
        \label{fig:C}
\end{subfigure}
\begin{subfigure}[t]{0.12\textwidth}
    \includegraphics[width=1\textwidth]{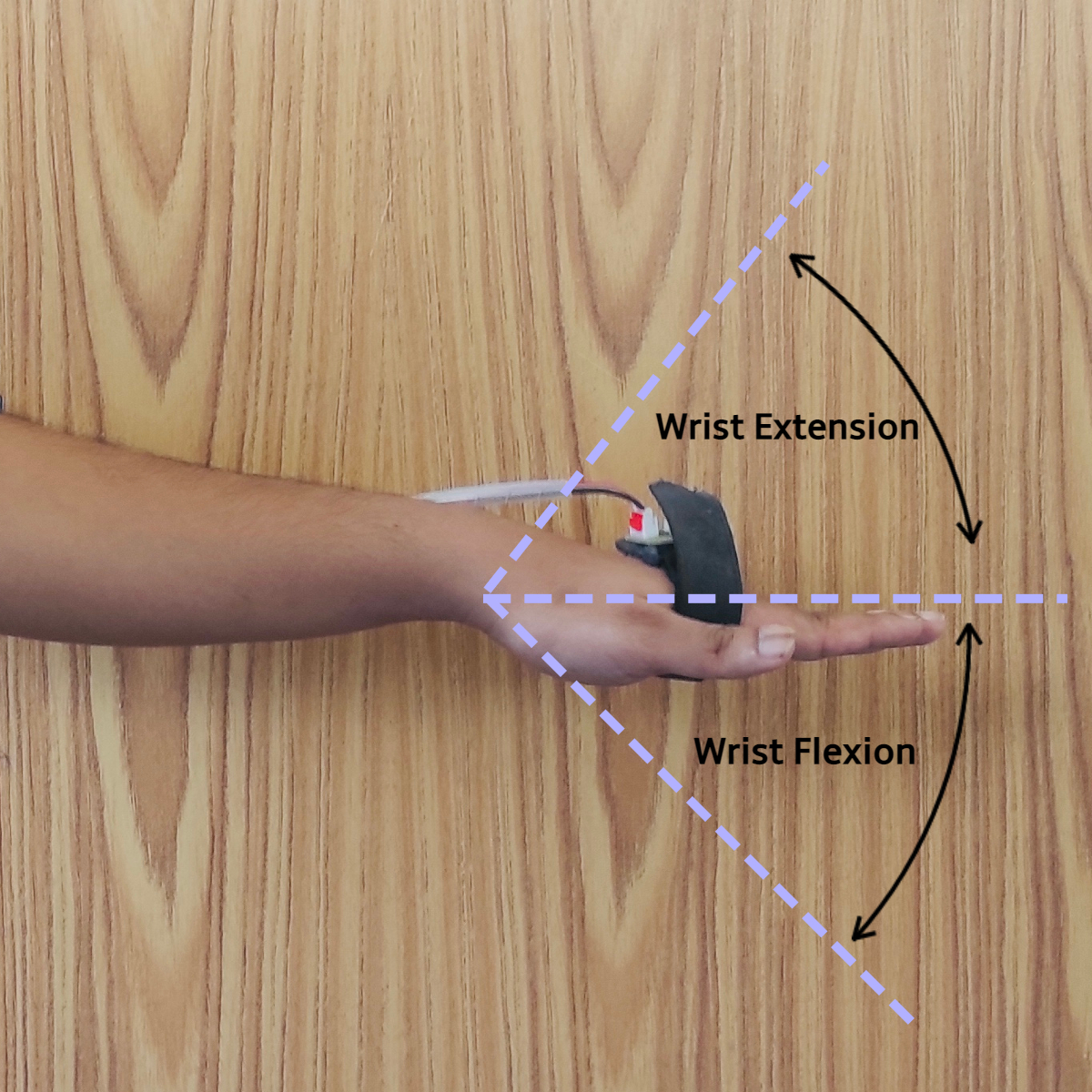} 
        \caption{Wrist Flexion (H4) and Extension (H5) }
        \label{fig:D}
\end{subfigure}
\hspace{0.5cm}
\begin{subfigure}[t]{0.12\textwidth}
    \includegraphics[width=1\textwidth]{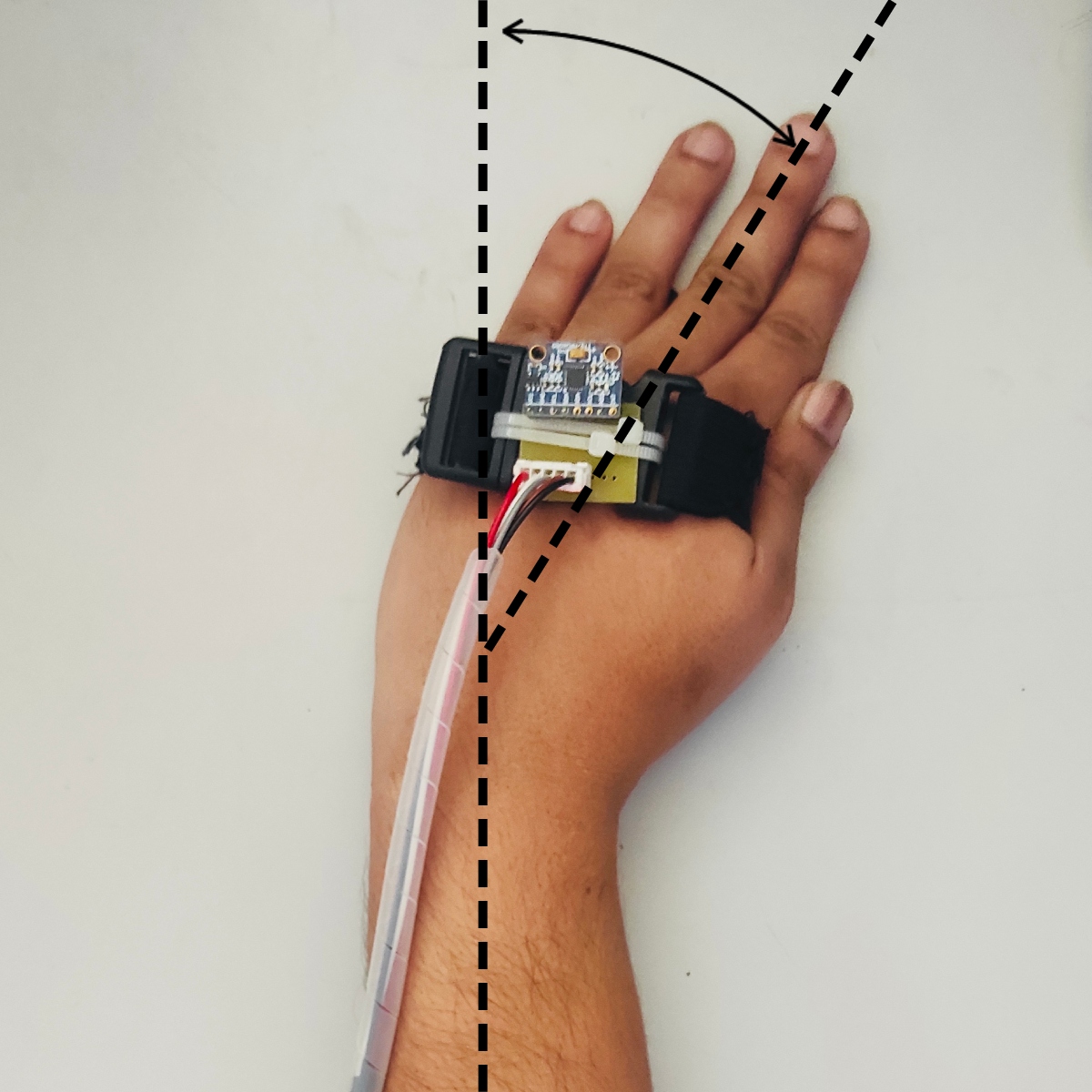} 
        \caption{Wrist Ulnar (H6)}
        \label{fig:E}
\end{subfigure}
\hspace{0.5cm}
\begin{subfigure}[t]{0.12\textwidth}
    \includegraphics[width=1\textwidth]{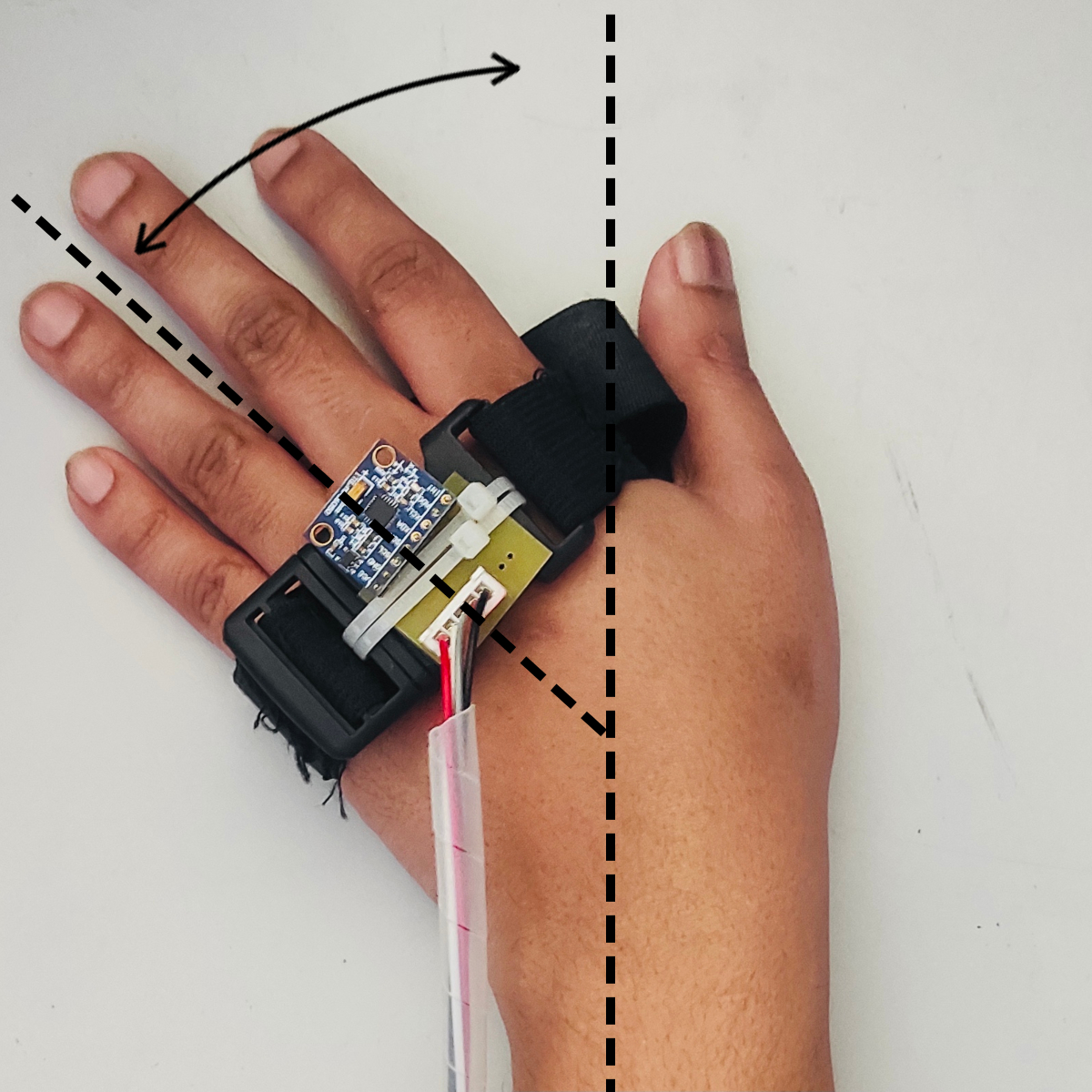} 
        \caption{Wrist Radial (H7)}
        \label{fig:F}
\end{subfigure}

\begin{subfigure}[t]{0.2\textwidth}
    \includegraphics[width=1\textwidth]{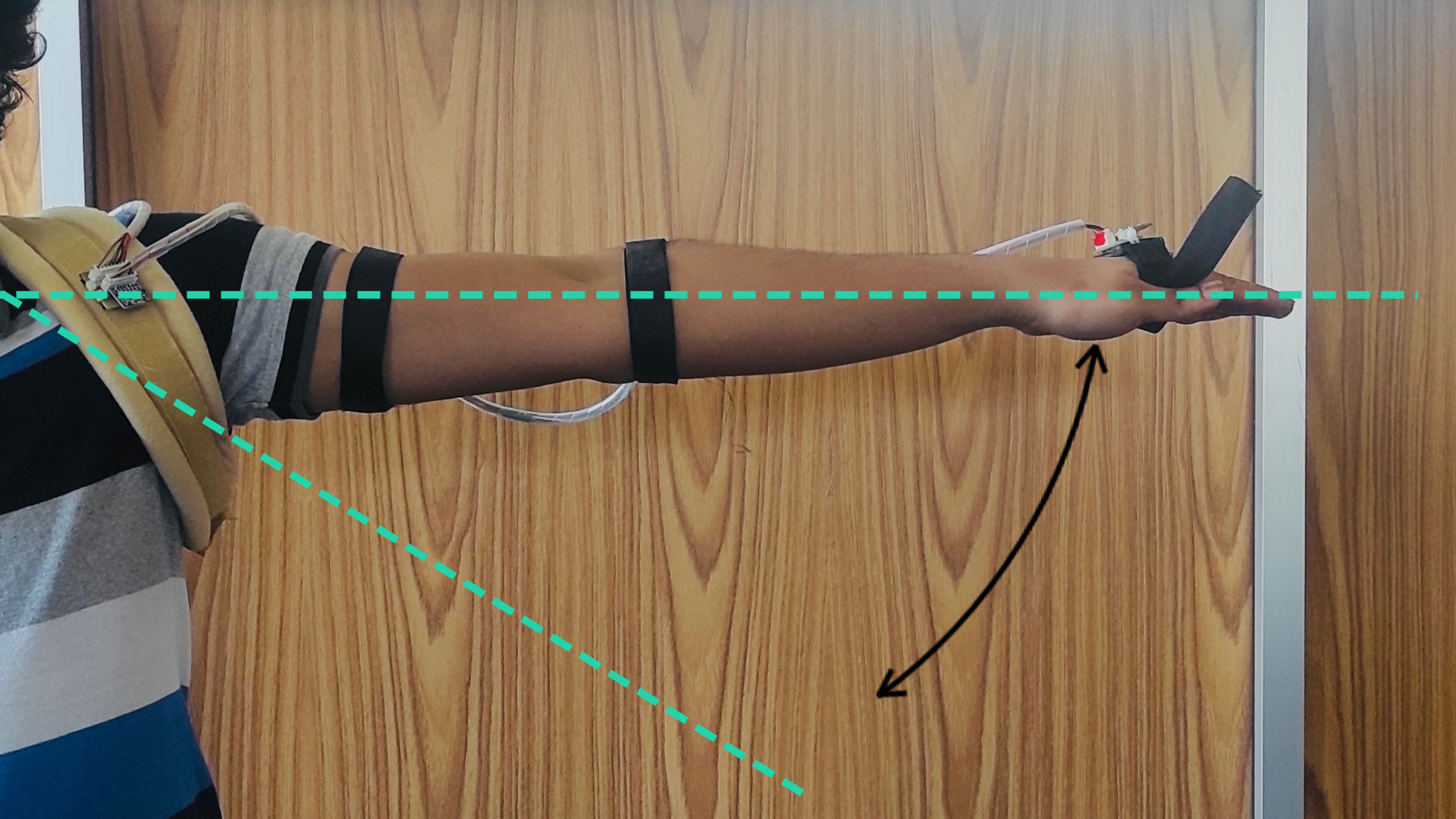} 
        \caption{Shoulder Abduction Adduction (H8)}
        \label{fig:G}
\end{subfigure}
\begin{subfigure}[t]{0.2\textwidth}
    \includegraphics[width=1\textwidth]{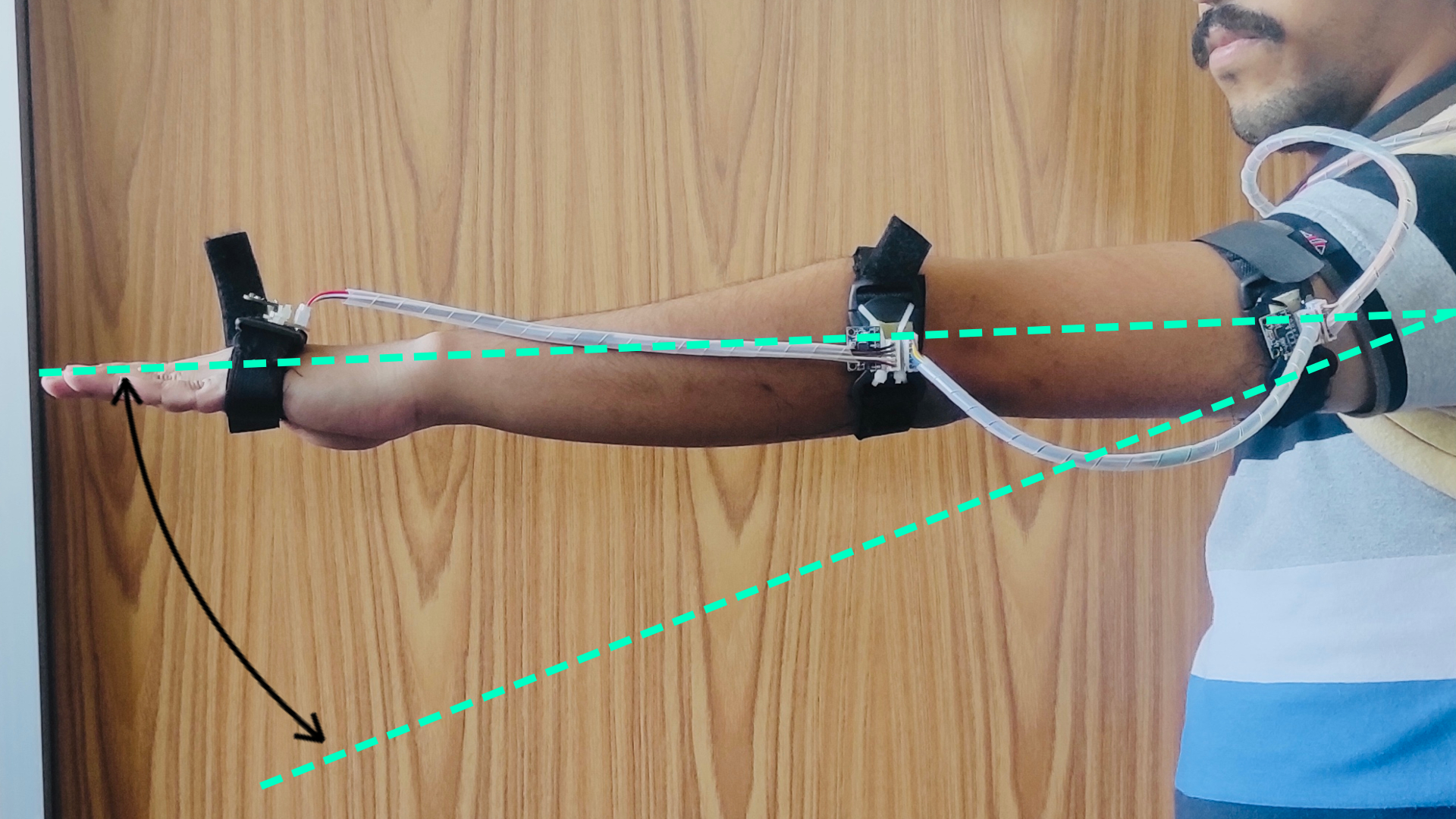} 
        \caption{Shoulder Flexion and Extension (H9)}
        \label{fig:H}
\end{subfigure}
\begin{subfigure}[t]{0.2\textwidth}
    \includegraphics[width=1\textwidth]{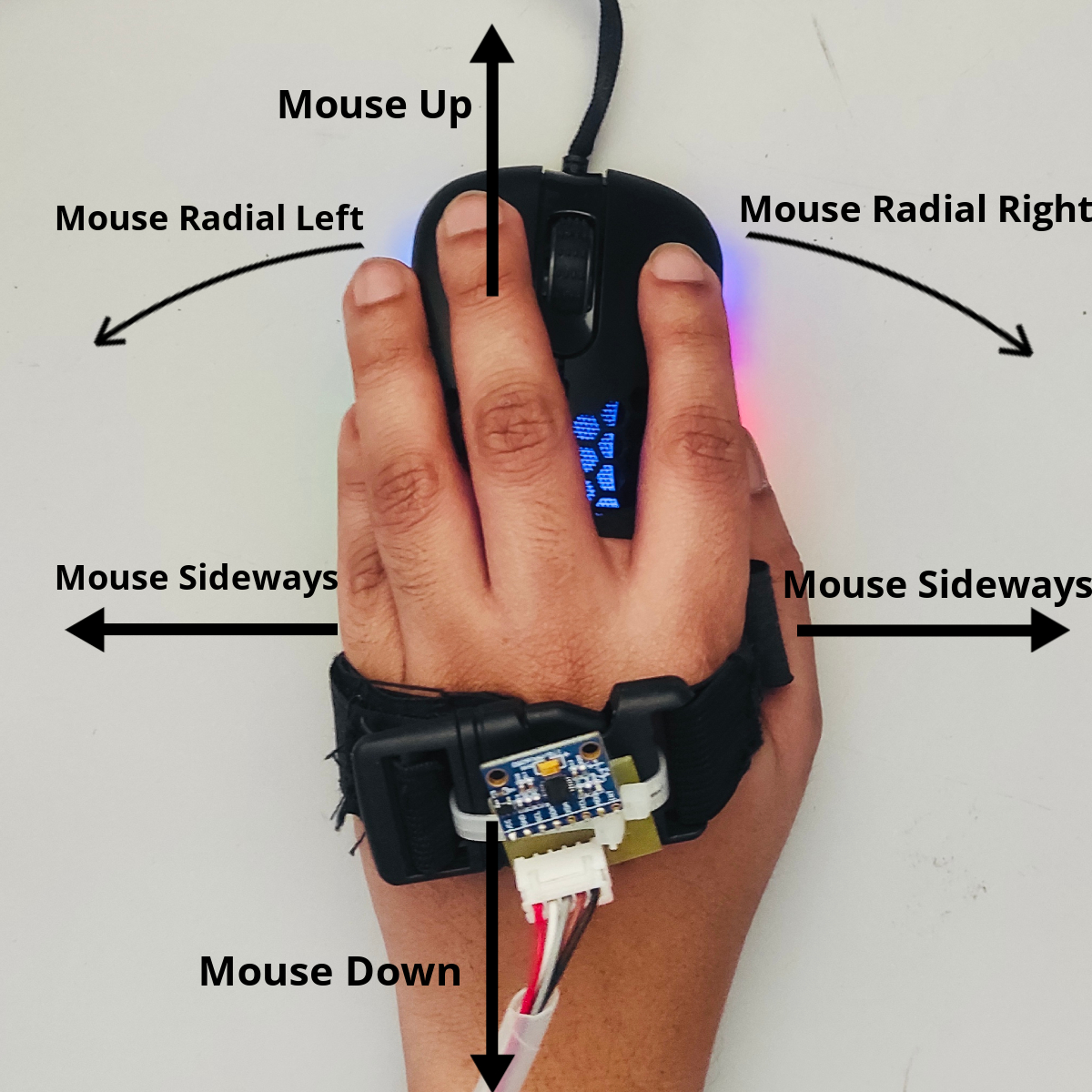} 
        \caption{Mouse Movements (H10, H11, H12, H13)}
        \label{fig:I}
\end{subfigure}
\caption{Hand Movements}
\label{fig:hand_mov}
\end{figure}

\begin{figure}[!t]
\centering

\begin{subfigure}[t]{0.35\textwidth}
    \includegraphics[width=1\textwidth]{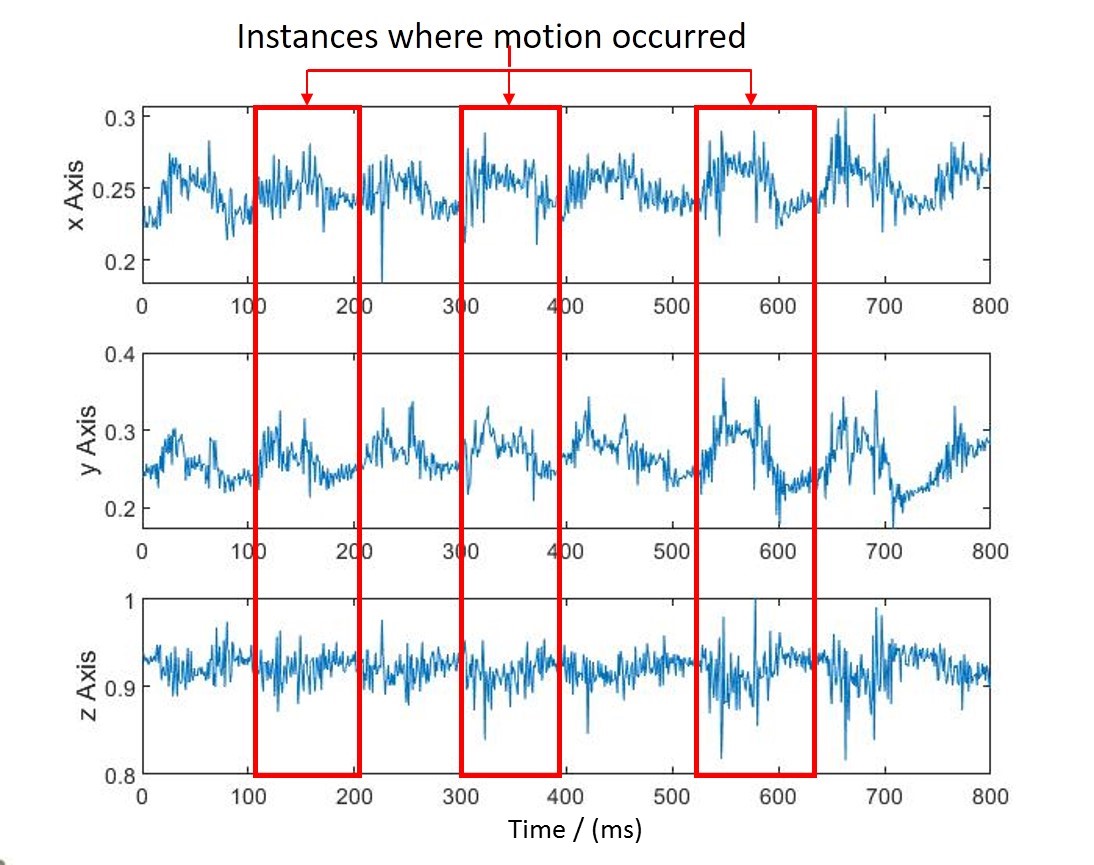}
    %\resizebox{0.5\linewidth}{!}{\input{figures/handaccdata.jpg}}  
        \caption{Accelerometer Data}
        \label{fig:A}
\end{subfigure}
\begin{subfigure}[t]{0.35\textwidth}
    \includegraphics[width=1\textwidth]{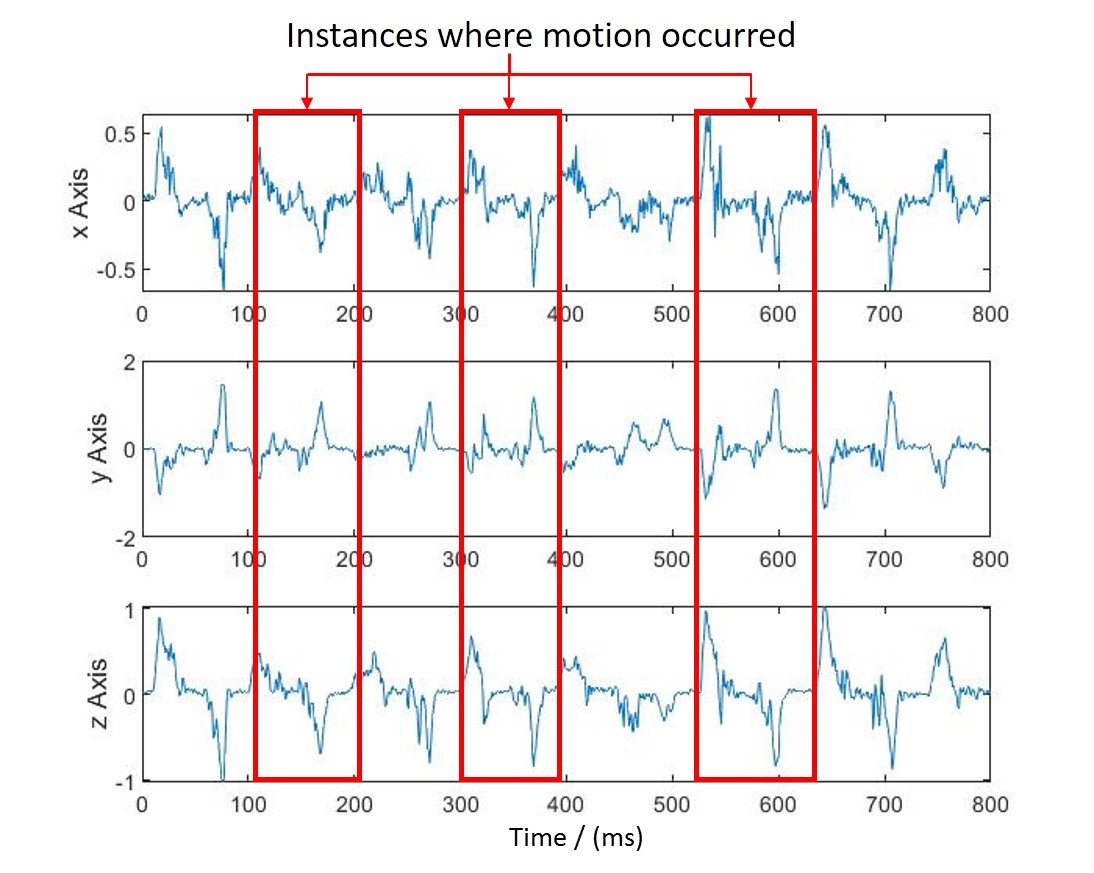}
    %\resizebox{0.5\linewidth}{!}{\input{figures/Pictyure2}}  
        \caption{Gyroscope Data}
        \label{fig:B}
\end{subfigure}
\caption{AS1 Sensor Readings for Power Grip Movement}
\label{fig:power_grip}
\end{figure}

\begin{figure}[!t]
\centering

\begin{subfigure}[t]{0.15\textwidth}
    \includegraphics[width=1\textwidth]{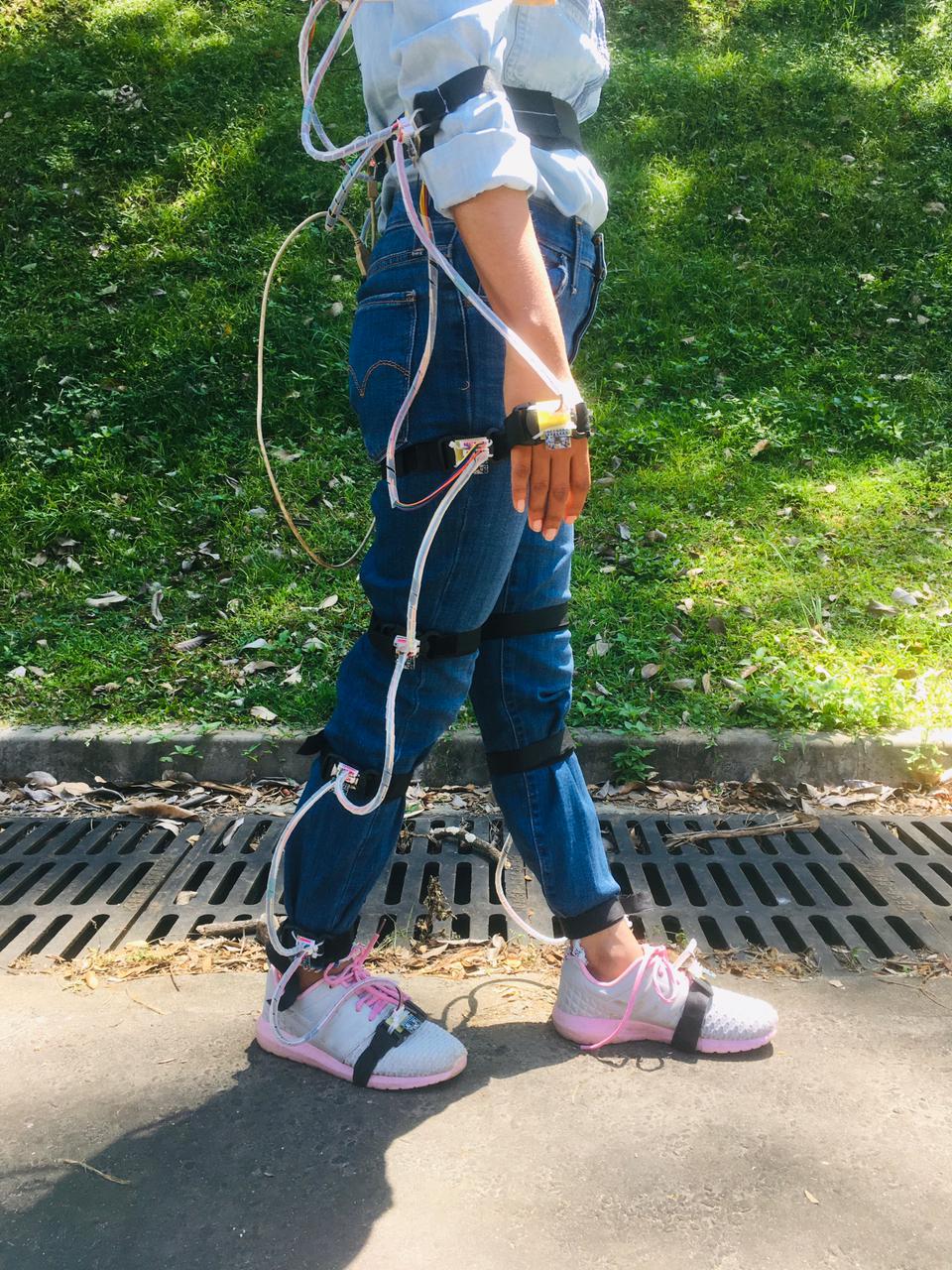}
        \caption{Plain Terrain}
        \label{fig:A}
\end{subfigure}
\begin{subfigure}[t]{0.15\textwidth}
    \includegraphics[width=1\textwidth]{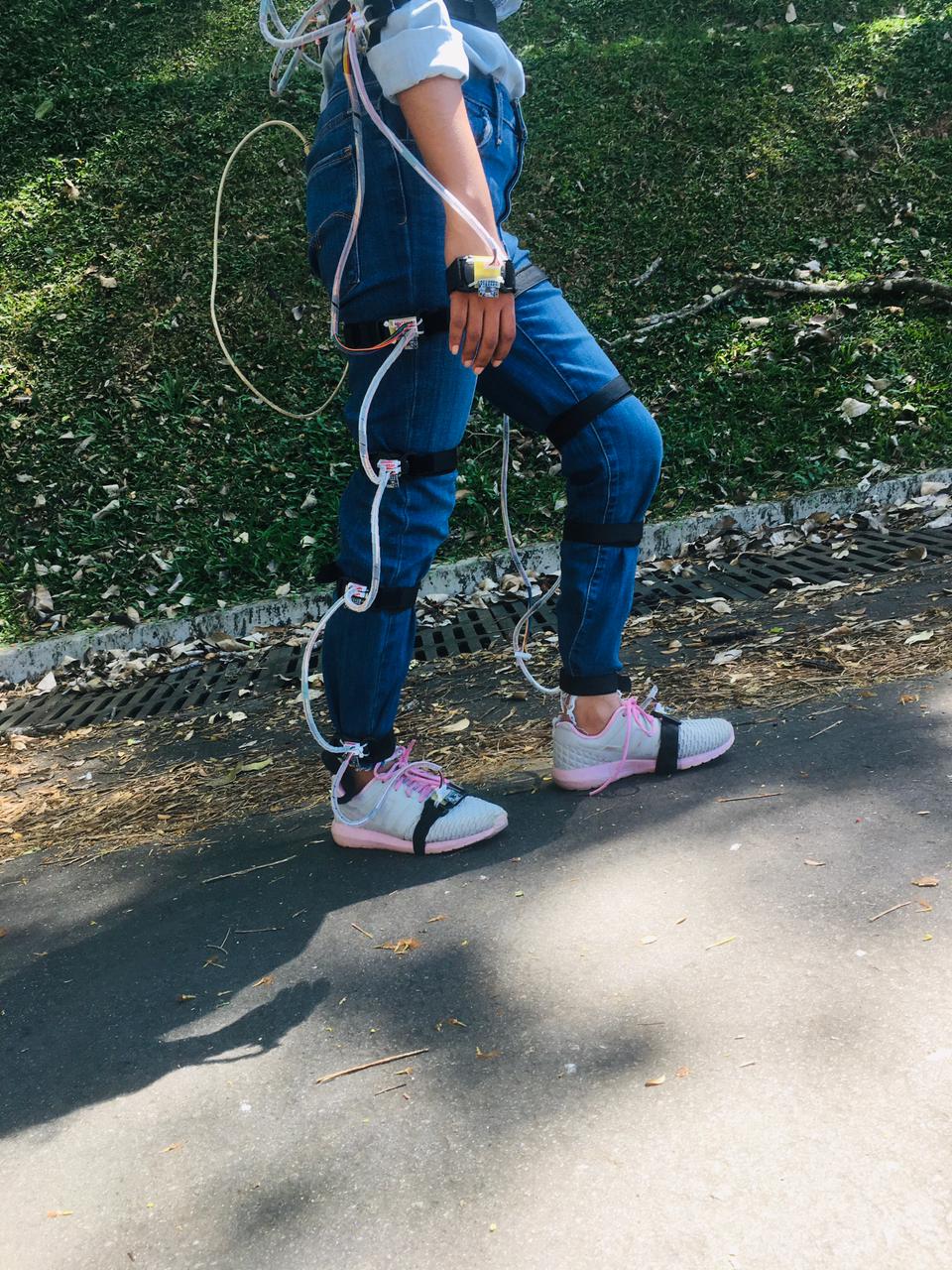} 
        \caption{Steep Terrain}
        \label{fig:B}
\end{subfigure}
\begin{subfigure}[t]{0.15\textwidth}
    \includegraphics[width=1\textwidth]{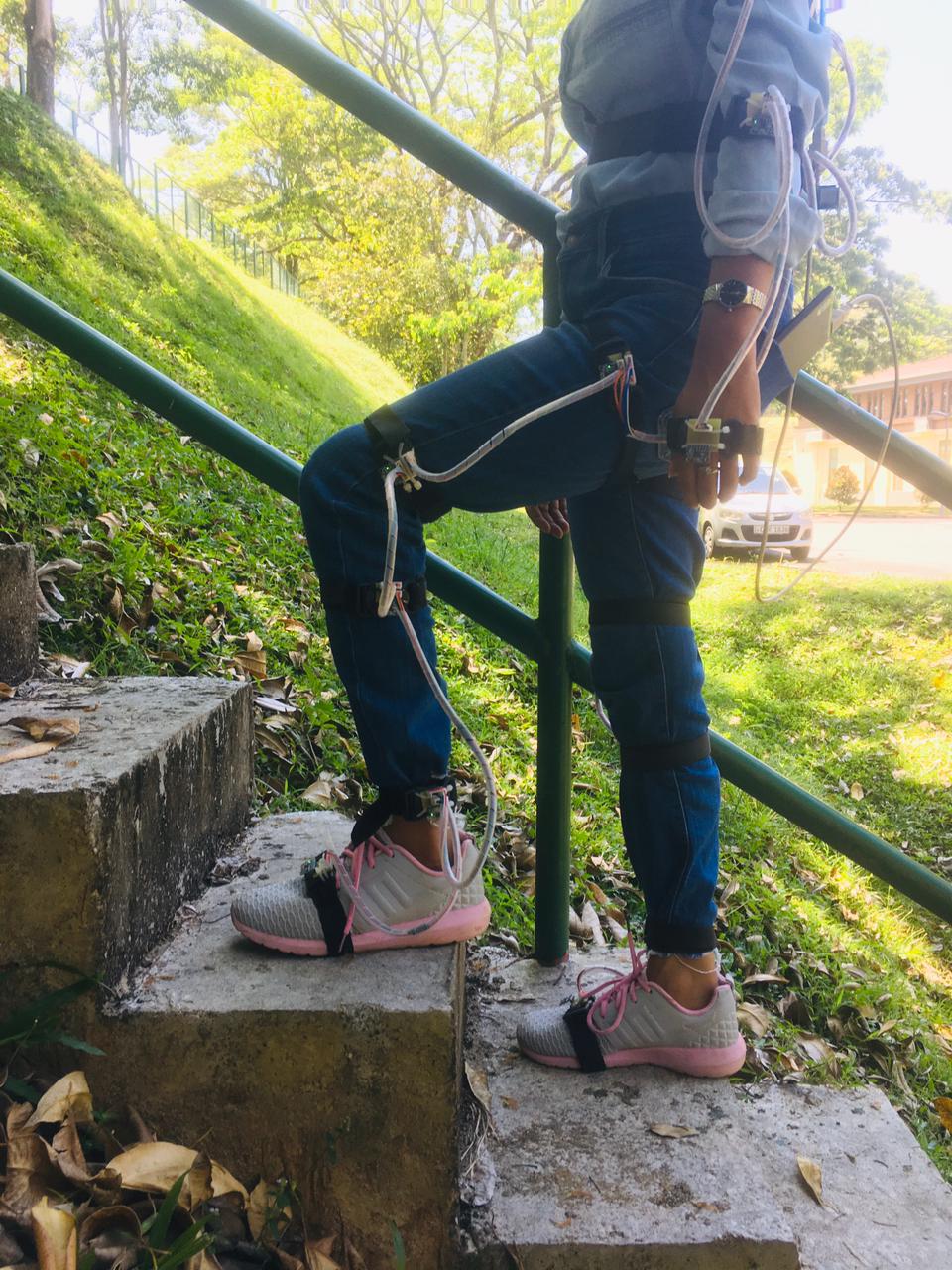} 
        \caption{Staircase}
        \label{fig:C}
\end{subfigure}
\caption{Walking Terrains with Sport Shoes}
\label{fig:walk_image}
\end{figure}

\begin{figure}[!t]
     \centering
     \begin{subfigure}[b]{0.15\textwidth}
         \centering
         \includegraphics[width=25mm]{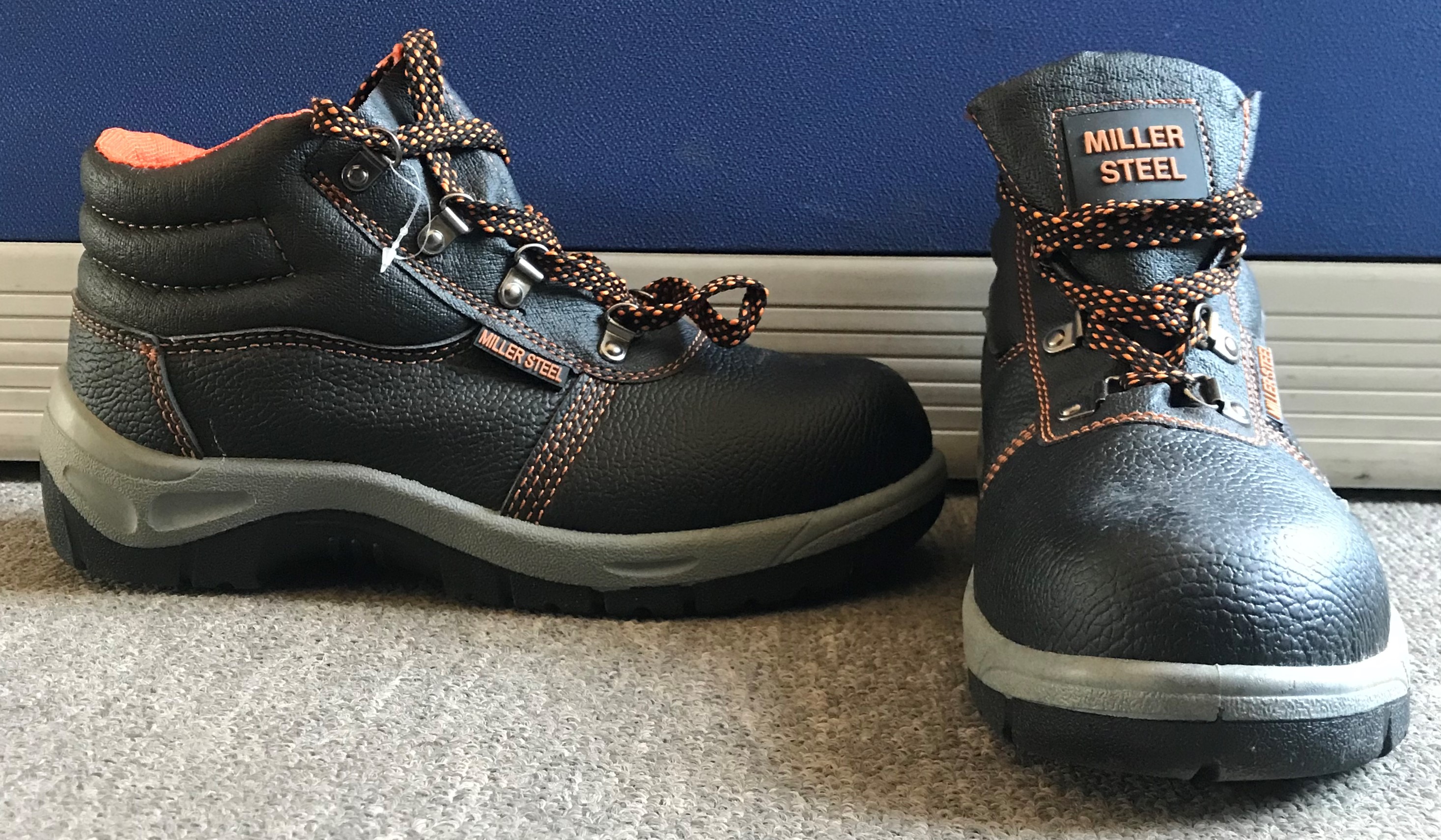}
         \caption{Safety Boots}
         \label{fig:boots}
     \end{subfigure}
     \hfill
     \begin{subfigure}[b]{0.15\textwidth}
         \centering
         \includegraphics[width=29mm]{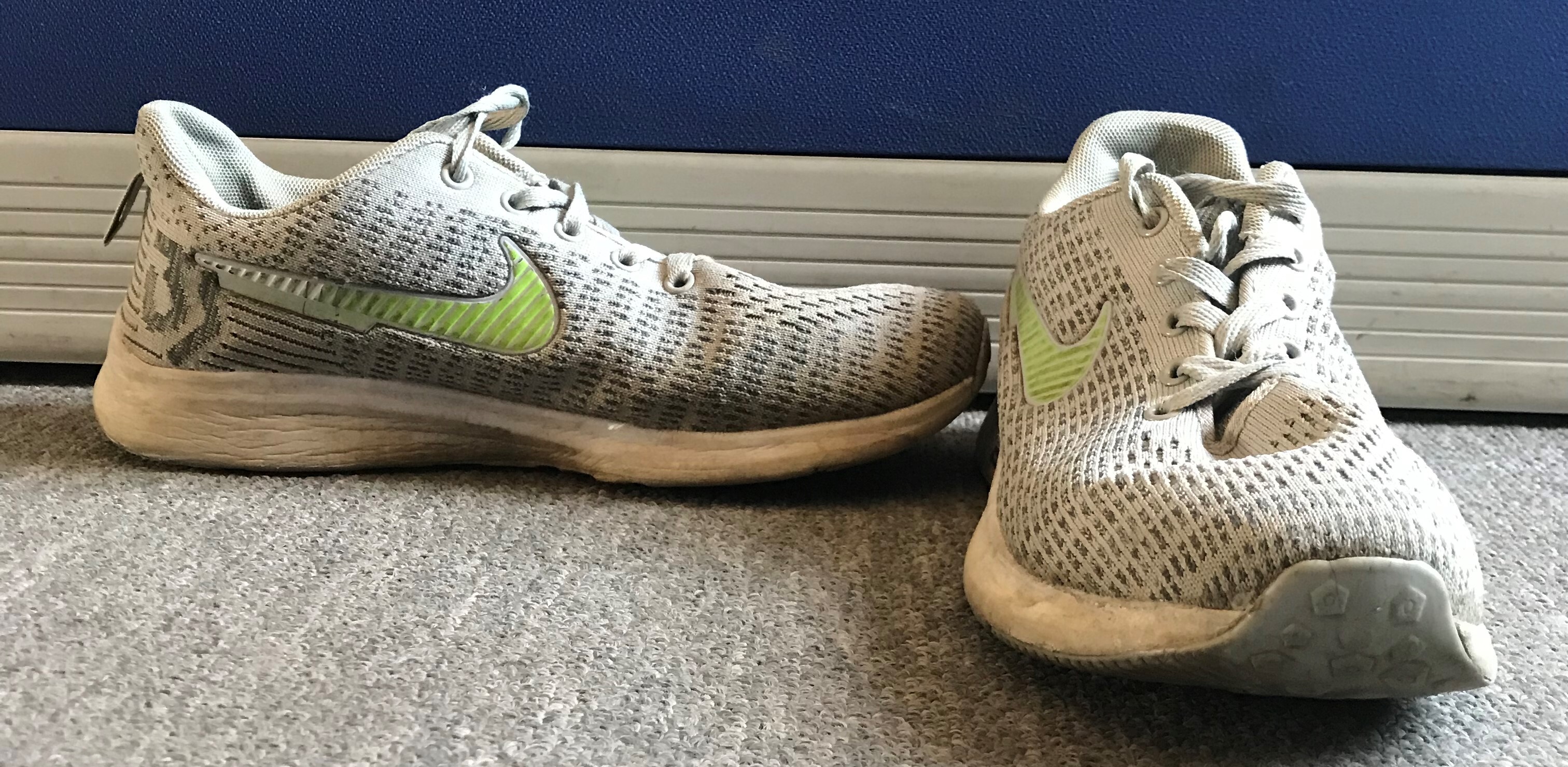}
         \caption{Sport Shoes}
         \label{fig:deks}
     \end{subfigure}
     \hfill
     \begin{subfigure}[b]{0.15\textwidth}
         \centering
         \includegraphics[width=27mm]{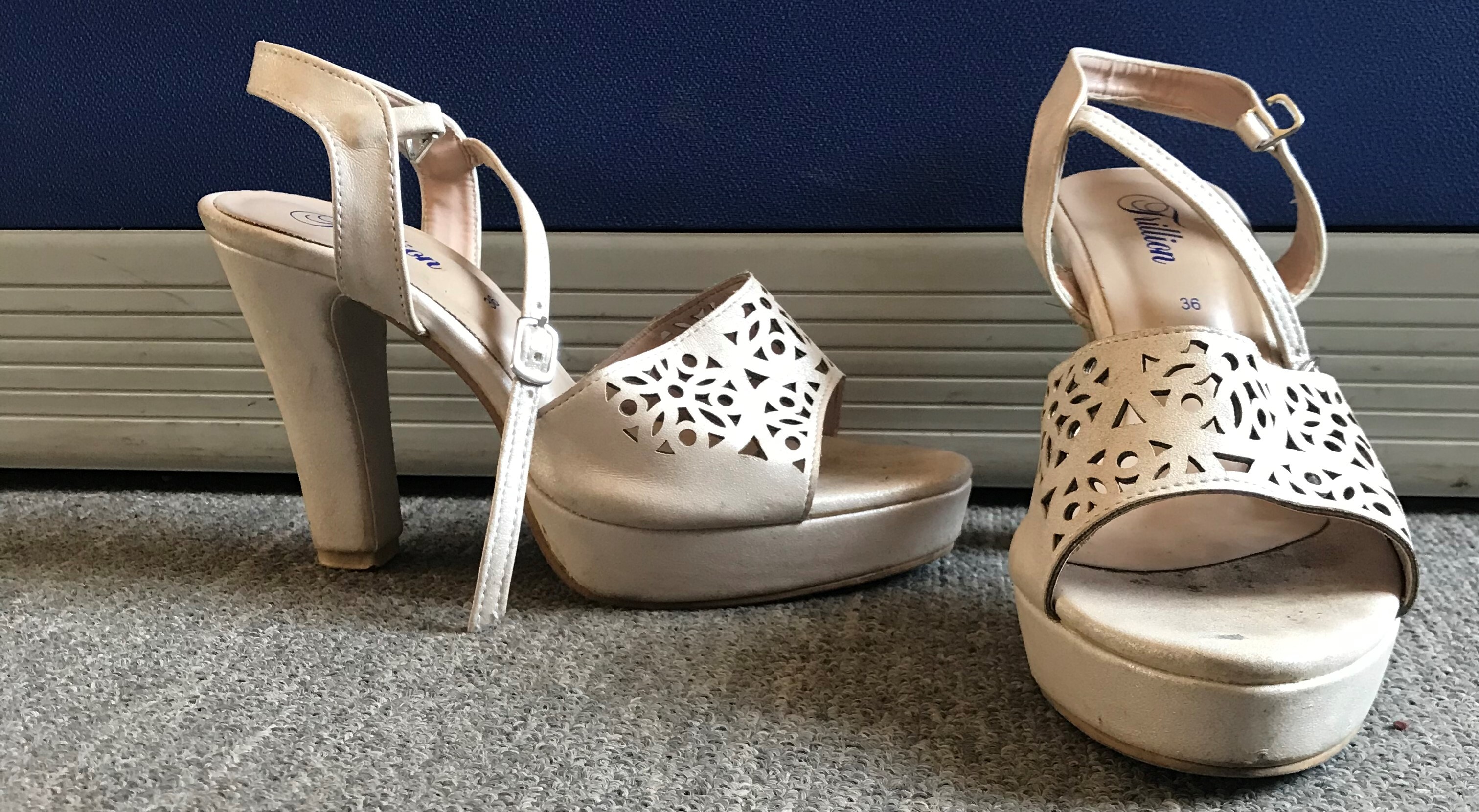}
         \caption{High Heels}
         \label{fig:heels}
     \end{subfigure}
        \caption{Three Shoe types}
        \label{fig:three graphs}
\end{figure}

\subsubsection{Functionality} 

The following list summarizes the functionality of each component:

\begin{itemize}
  \item Central Arduino – Acts as the central unit for synchronizing and controlling the other microcontrollers.
  \item  Module Arduino 1-4 – Collect data from the sensors and store them when prompted by the central unit.
  \item RTC – Real-time clock for date and time stamping of the data collected.
  \item Module SD 1-4 – To save the accelerometer and gyroscope data of each module.
  \item Start /Stop push button– To start and stop data capturing.
  \item Error push button - To mark abnormal occurrences while collecting data
\end{itemize}

\subsection{Measurement}

%\begin{table}[!t]
%    \centering
%    \caption{Limb Movements}
%    \label{tab:movements}
%    \begin{tabular}{cc}
%    \hline
%      Hand Movements & Walking Patterns \\
%       \hline\hline
%        Elbow Flexion and Extension & on a plain with Sports-shoes\\
%        Elbow Pronation and Supination & on a plain with Safety Boots \\
%        Mouse radial top left & on a plain with High Heels\\
%        Mouse radial top right & up a staircase with Sports-shoes\\
%        Mouse sideways & up a staircase with Safety Boots\\
%        Mouse up-down & up a staircase with High Heels\\
%        Power grip & down a staircase with Sports-shoes\\
%        Shoulder Abduction and Adduction & down a staircase with Safety Boots\\
%        Shoulder Flexion and Extension & down a staircase with High Heels\\
%        Wrist Flexion  & up a slope with Sports-shoes \\
%        Wrist Ulnar &  up a slope with Safety Boots\\
%        Wrist Radial & up a slope with High Heels \\
%        Wrist Extension & down a slope with Sports-shoes\\
%         & down a slope with Safety Boots\\
%          & down a slope with High Heels\\

%        \hline
           
%    \end{tabular}    
% \end{table}

\begin{table}[!t]
 \setlength\extrarowheight{2pt}
 \centering
 \caption{Hand Movements}
 \label{tab:hand_movements}
\begin{tabular}{|p{0.1cm}|p{0.01cm}|p{0.05cm}|p{0.05cm}|p{0.05cm}|p{0.05cm}|p{0.05cm}|p{0.05cm}|p{0.05cm}|p{0.05cm}|p{0.05cm}|p{0.1cm}|p{0.1cm}|p{0.2cm}|p{0.1cm}|}
\cline{3-15}
  \multicolumn{2}{c|}{\multirow{2}{*}{}} & \multicolumn{9}{c|}{Movement} & \multicolumn{4}{c|}{Mouse Movement} \\ \cline {3-15}
  \multicolumn{2}{c|}{} & \rotatebox{90}{Flexion\phantom{.}} & \rotatebox{90}{Extension\phantom{.}} & \rotatebox{90}{Pronation\phantom{.}} & \rotatebox{90}{Supination\phantom{.}} & \rotatebox{90}{Abduction\phantom{.}} & \rotatebox{90}{Adduction\phantom{.}} & \rotatebox{90}{Ulnar\phantom{.}} & \rotatebox{90}{Radial\phantom{.}} & \rotatebox{90}{Power Grip\phantom{.}} & \rotatebox{90}{Radial Left\phantom{.}} & \rotatebox{90}{Radial Right\phantom{.}} & \rotatebox{90}{Up-Down\phantom{.}} & \rotatebox{90}{Sideways\phantom{.}} \\ \hline

  \multirow{3}{*}{\rotatebox{90}{{Rotational\phantom{.}}}} & \multicolumn{1}{c|}{Wrist}  & \checkmark \newline \tiny{H4} & \checkmark \newline \tiny{H5} & & & & & \checkmark \newline \tiny{H6} & \checkmark \newline \tiny{H7} & \checkmark \newline \tiny{H3} & \checkmark \newline \tiny{H10} & \checkmark \newline \tiny{H11} & &   \\  \cline{2-15} 
  
 & \multicolumn{1}{c|}{Elbow} & \multicolumn{2}{c|}{\checkmark \newline \tiny{H1}}  & \multicolumn{2}{c|}{\checkmark \newline \tiny{H2}} & & & & & & & & &   \\\cline{2-15}

   & \multicolumn{1}{c|}{Shoulder} & \multicolumn{2}{c|}{\checkmark \newline \tiny{H9}} & & & \multicolumn{2}{c|}{\checkmark \newline \tiny{H8}} & & & & & & & \\ \hline

    \multicolumn{2}{|c|}{Linear}  & & & & & & & & & & & & \checkmark \newline \tiny{H12} & \checkmark \newline \tiny{H13}  \\ \hline

\end{tabular}

\end{table}

\begin{table}[!t]
    \centering
    \caption{Types of Shoes}
    \label{tab:shoes}
    \begin{tabular}{ccc}
    \hline
       Shoe & Height change & Weight per leg \\
       \hline\hline
        Sport Shoe & 2.0cm & 200 g \\
        High Heels & 10.0cm & 225 g \\
        Safety Boots & 2.5cm & 575 g \\
        \hline
           
    \end{tabular}    
\end{table}

\begin{table}[!t]
 \setlength\extrarowheight{2pt}
 \centering
 \caption{Walking Patterns}
 \label{tab:Walking_Patterns}
\begin{tabular}{c|c|c|c|c|}
\hline
  \multicolumn{2}{|c|}{\multirow{2}{*}{Terrain}} & \multicolumn{3}{c|}{Shoe Type} \\ \cline{3-5}
  \multicolumn{1}{|c}{} & & Sports Shoe & Working Boots & Heels  \\   \hline
 \multicolumn{2}{|c|}{Plain} & \checkmark & \checkmark & \checkmark \\ \hline
 \multicolumn{1}{|c|}{\multirow{2}{*}{Staircase}} & Up & \checkmark & \checkmark & \checkmark \\  \cline{2-5}
\multicolumn{1}{|c|}{} & Down & \checkmark & \checkmark & \checkmark\\  \cline{1-5}
\multicolumn{1}{|c|}{\multirow{2}{*}{Steep}} & Up & \checkmark & \checkmark & \checkmark\\  \cline{2-5}
\multicolumn{1}{|c|}{} & Down & \checkmark & \checkmark & \checkmark \\  \cline{1-5}
\end{tabular}
\end{table}

\subsubsection{ Hand}

A single unit was used to take readings and the subject was advised to perform activities repeatedly/periodically. This study focused on classifying 13 hand movements as described in Table \ref{tab:hand_movements} and in Figure \ref{fig:hand_mov}. H1, H2, H8, and H9 movements are combined movements that are represented by merged cells in Table \ref{tab:hand_movements}. Furthermore, Figure \ref{fig:power_grip} shows the readings obtained from the smallest movement among the 13, where only sensor (AS1) was triggered.

\subsubsection{Walking Pattern}
To properly capture the versatile nature of the working environment, it was decided to use three different terrains to walk on and
collect data. A steep slope, a staircase (with
at least ten steps), and a planar surface were used as
the three terrains which are demonstrated in Figure \ref{fig:walk_image}. It was observed that a person uses two different walking styles while walking upwards and downwards along a staircase or a slope. According to these observations, those two were treated as two different walking patterns. 
The weight and the height of the heel of a shoe have a great impact on the walking pattern \cite{heel1, heel2}. This impact was captured using three types of shoes, as mentioned in Table \ref{tab:shoes} and Figure \ref{fig:three graphs}. A light weighted, seat-heeled sports shoe was used as the reference shoe. Safety boots with a heavy tow covering were used to mimic heavy shoes while a pair of high heels was used to increase the heel height with respect to the toes. Table \ref{tab:Walking_Patterns} represents the 15 walking patterns that were analyzed in this study. The second case study as supposed to the first, focused on using specific limb movement patterns. Here the focus was on the task of walking and how it gets affected due to external environmental and attire changes. 

%\subsection{Observations}
%The sensor placement on the limbs has permitted to capture of a wide range of motions for both hand and leg movements. It was evident from the number of triggered sensors during the hand movements, that the smallest (Power Grip) to the largest hand movement (shoulder pivoted movements) could be captured from the designed device. Figure \ref{fig:power_grip} shows the AS1 sensor reading of the hand for the 'Power Grip' movement. 

\subsection{Data conditioning through active sensor selection}

Sensor selection would be different for different applications. For instance, the sensors that are dropped for a certain movement pattern due to signal power might be critically important for another movement. It enables active sensor selection based on signal power. In addition to that, if two sensors' motion patterns are highly correlated, and one does not convey any additional information compared to the other that sensor's information is refunded. Hence the below-mentioned stage I and II were motivated based on the above considerations.  
For the hand movements, four IMU sensors are used, each with three directional measurements (four IMU with an accelerometer and a gyroscope, each sensor with three directional measures resulting in 24 overall sensor dimensions [$4\times2\times3 = 24$]). Similarly, for walking patterns 72 sensor dimensions were acquired. ([$12\times2\times3 = 72$]). Now let us consider three filtering stages and 2-dimensional reduction stages that were used to obtain effective and efficient data representation.

\subsubsection{\textbf{Stage I (S1)} - Signal Power based filtering}
As a result of empirical observation, some directional sensor data exhibit fewer power fluctuations than others (let us consider $p$ number of sensors had less than $\gamma$, average power), resulting in them being considered lesser contributors for that specific movement hence they were excluded from further processing. At this point, the number of sensor measurements that contribute to the algorithm would be $N - p$ ($N$ - Number of sensor dimensions)

\subsubsection{\textbf{Stage II (S2)} - Clustering measurement based filtering}
The sensor-wise inter-cluster and intra-cluster distances \cite{intra_inter1} are studied to explore the influence of directional sensor readings on identifying various movements, where a cluster is defined as a hand movement. For example,  H1 - H13 are considered different clusters. A directional sensor measurement with less intra-cluster distance and more inter-cluster distance is a desirable directional sensor measurement to employ while clustering. This study focuses on the ratio of intra-cluster length to inter-class distance to combine both elements. If the ratio is lower, the directional-sensor measurement contributes more to the clustering process. After filtering out the most desirable directional-sensor measures ($r$) with a suitable threshold for the cluster distance ratio (now the $N - p$ dimension space is reduced to $r$), the measurements are concatenated to make it more realistic because the entire hand movement comprises all of the sensor readings. The data points were then flattened while maintaining the temporal domain, and classic machine learning techniques were performed using supervised learning.

\subsubsection{\textbf{Stage III (S3)} - Data smoothing and representation through the Kalman Filter}
The data from the accelerometers and the gyroscopes of the device contains noisy information about the angular rotations and inertial forces acting on the sensors. Hence, these sensors in turn are indicators of the angular velocity and the accelerations of the limbs. 

The device is expected to categorize different movement patterns of the human body. Therefore, rather than considering using the data directly to categorize the movements, the data can be transformed into a form that is a better representation of the movements.

The gyroscope of the sensors measures the angular velocity of the limb to which it is attached. Also, the accelerometer measures the acceleration of the limb as well as the gravitational force. When considered in their raw form, the method that is used to categorize human motion with this information would have a better performance if they were converted into another form that is a better indicator of the limb orientations. 

The gyroscope and the accelerometer data can be used to obtain the orientation of each limb which can then in turn be used to find the orientation of each limb relative to each other and reconstruct the motion.

%\subsubsection{Use of the Kalman filter}

The direct way to obtain the orientation of a gyroscope is by integrating the angular velocity over time to obtain the angular position of the gyroscope. That can only be used for a short time period due to the inclusion of gyroscopic drift. Similarly, when the accelerometer data are used independently, the estimations are only valid when the accelerometers are at rest. Therefore, as stated in \cite{kalman_for_attitude}, the orientation can be obtained using both the gyroscope and accelerometer data using sensor fusion with an extended Kalman filter with a non-linear estimation model to get the most accurate estimate possible. \Cref{eq:state_vector,eq:error_state,eq:bias_prop,eq:omega_prop,eq:covaiance_est,eq:measurement_residual,eq:covariance,eq:Kalman_gain,eq:state_update,eq:bias_update,eq:omega_update,eq:quaternion_update,eq:covarience_update} represents the Kalman filter equations used in the pre-filtering process. 

With multiple sensors, the sensor orientation calculation with rotational or transformation matrices will be computationally costly to execute. To reduce the extra burden of computational algorithms, the quaternion is calculated for a sensor. The results will be then used as inputs for the Kalman filter, which will then provide a $4 \times n$ quaternion matrix as the output for a data point where $n$ is the number of sensors.   
\\

\noindent State Equations
\begin{equation}
   \label{eq:state_vector}
   x_k=\begin{bmatrix}
       \hat{q}_k\\\hat{b}_k
   \end{bmatrix}
\end{equation}
\begin{equation}
    \label{eq:error_state}
    \Tilde{x}_k =\begin{bmatrix}
        \delta\theta_k\\{\Delta}\hat{b}_k
    \end{bmatrix}
\end{equation}
Propagation
\begin{equation}
    \label{eq:bias_prop}
    \hat{b}_{k|k-1} = \hat{b}_{k-1|k-1}
\end{equation}
\begin{equation}
    \label{eq:omega_prop}
    \hat{\omega}_{k|k-1}=\omega_{m_k}-\hat{b}_{k|k-1}
\end{equation}
\begin{equation}
    \label{eq:covaiance_est}
    P_{k|k-1}=F_kP_{k|k-1}F^T_k+Q_k
\end{equation}
Update
\begin{equation}
    \label{eq:measurement_residual}
    \Tilde{r}_{k|k-1}=z_k-\hat{z}_{k|k-1}
\end{equation}
\begin{equation}
    \label{eq:covariance}
    S_k=H_kP_{k|k-1}H^T_k+R_k
\end{equation}
\begin{equation}
    \label{eq:Kalman_gain}
    K_k=P_{k|k-1}H^T_kS^{-1}_k
\end{equation}
\begin{equation}
    \label{eq:state_update}
    \Delta\hat{x}_k = \begin{bmatrix}
        \delta\hat{\theta}_k\\{\Delta}\hat{b}_k
    \end{bmatrix} = \begin{bmatrix}
         2\delta\hat{q}_k\\{\Delta}\hat{b}_k
    \end{bmatrix}=K_{k}\tilde{r}_{k|k-1}
\end{equation}
\begin{equation}
    \label{eq:bias_update}
    \hat{b}_{k|k} =\hat{b}_{k|k-1}+\Delta\hat{b}_k 
\end{equation}
\begin{equation}
    \label{eq:omega_update}
    \hat{\omega}_{k|k} = \omega_{m_k}-\hat{b}_{k|k}
\end{equation}
\begin{equation}
    \label{eq:quaternion_update}
    \hat{q}_{k|k}=\frac{\delta\hat{q}_k}{||\delta\hat{q}_k||}\otimes\hat{q}_k
\end{equation}
\begin{equation}
    \label{eq:covarience_update}
    P_{k|k} = (I-K_kH_k)P_{k|k-1}
\end{equation}

\begin{table}[t]
    \centering
    \caption{The Terminologies}
    \label{tab:kalman_filter_term}
    \begin{tabular}{cl}
    \hline
       Terminologies & Meanings \\
       \hline\hline
        $k$ & Time step\\
        $x_k$ & State Vector \\
        $\tilde{x}_k$ & Error Vector \\
        $\hat{b}_k$ & Gyro Bias \\
        $\delta\theta_k$ & Error Angle Vector \\
        ${\Delta}\hat{b}_k$ & Gyro Bias Error \\
        $\hat{\omega}_k$ & Estimated Rotational Velocity \\
        ${\omega}_{m_{_k}}$ & Measured Rotational Velocity \\
        $P_{k|k-1}$ & State covariance matrix \\
        $F_k$ & State Transition Matrix \\
        $Q_k$ & Noise Covariance matrix \\
        $\Tilde{r}_{k|k-1}$ & Measurement Error \\
        $z_k$ & Actual Measurement \\
        $\hat{z}_{k|k-1}$ & Measurement Estimation\\
        $S_k$ & Residual Covariance Matrix \\
        $R_k$ & Measurement Error Covariance \\
        $H_k$ & Measurement Matirx \\
        $K_k$ & Kalman Gain Matrix \\
        $\Delta\hat{x}_{k}$ & State Correction Matrix \\
        $\hat{q}_{k|k}$ & Quaternion \\
        $\Delta\hat{b}_k$ & Bias error\\

    \end{tabular}    
\end{table}
\subsubsection{\textbf{Stage IV (S4)} - Feature space construction and dimension reduction through distance measures}

A distance measure is a comparative representation of a particular signal to a reference signal. The reference signal is selected in such a signal closer to all other signals in their cluster. In this study five distance measurements: Euclidean distance (ED), mean absolute percentage error (MAPE), Pearson correlation coefficient (PCC), Dynamic Time Warping (DTW), and Fréchet distance are used as they all concatenate the feature space in the time domain into a single point. For each data point, $n$ (number of classes) measurements are calculated for a single distance measurement with respect to $n$ reference hand movement signals which transforms 5$n$ new feature domains for a data point ($n$ reference signals x 5 distance measurements). 

For example, let us consider the classification of hand movements. Here, $n$ = 13, $N$ = 24, and let us assume $p$ as four, $r$ as 15, and 500-time samples per data point. After stage I, the tensor (dimension) that represents one data point is ($N$-$p$ = 24-4) 20. Then after stage II, the tensor of a data point is 15x500 (after concatenating and flattening this becomes 500). However, after stage II is followed by stage IV this tensor changes into 1x65 (5$n$ = 5x13) which is a 7.7-fold dimension reduction.    

\subsubsection{\textbf{Stage V (S5)} - PCA-based dimension reduction} After following the above stages Principle component analysis (PCA) was performed as the final dimensionality-reduction step with the intention of capturing the most important components in the data. 

\subsection{Machine Learning analysis for classification of the human motions}

\subsubsection{\textbf{Classification of hand movements}}

In terms of hand movement analysis, every movement has a similar average action period since the subject was told to remain seated and make hand motions at equal time intervals ($t$)  (500 ms is required to validate the units). 

The following combinations of stages are examined and Table \ref{tab:results_hand} summarizes them.

\begin{enumerate}

    \item Signal power-based, clustering measurement-based filtering, and PCA-based dimension reduction [S1, S2, S5].
    \item Signal power-based, clustering measurement-based filtering, and distance-measure-based, PCA-based dimension reduction [S1, S2, S4, S5].
    \item The Kalman Filter data smoother and PCA-based dimensional reduction [S3, S5].
    \item The Kalman Filter data smoother and distance measure-based, PCA-based dimensional reduction [S3, S4, S5].
    
\end{enumerate}

This study uses four supervised classifiers, namely Logistic Regression (LR), k-Nearest Neighbors (KNN), Random Forrest (RF), and Naive Bayes (NB), and compares the performance of the classified results.

\subsubsection{\textbf{Classification of walking patterns}}

Classification using the two best-performing combinations with (combination 2) and without (combination 4) the Kalman smoother is investigated after taking into account the outcomes of the classification of the hand movement. It is demonstrated in Table \ref{tab:results_walking} with the accuracies of the classification.

\begin{figure*}[!t]
\centering
\includegraphics[width=0.8\textwidth]{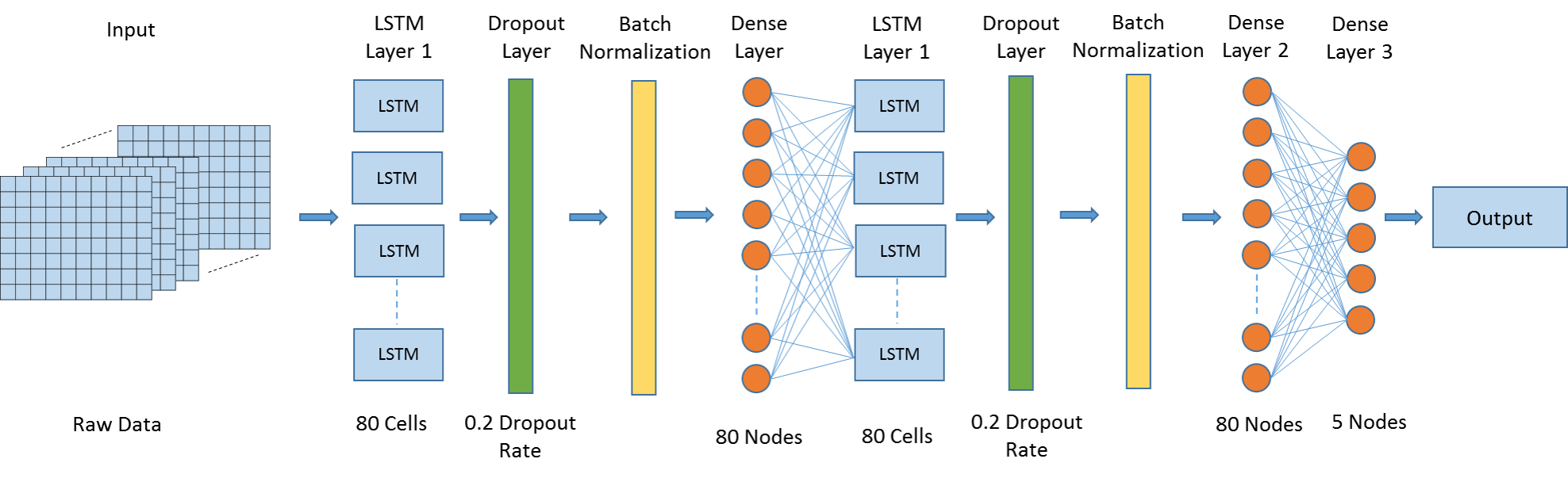}
\caption{LSTM Network Architecture}
\label{fig:lstm_code}
\end{figure*}

\subsection{Recurrent Neural Network analysis for human motion classification}
RNNs were designed to work with sequence prediction problems. Walking and hand gestures used to acquire data are sequential and repeated multiple times. Most of the natural moments are gradual and vary slightly over time. So RNN is the best suited for these types of sequence or pattern recognition. Due to the higher data acquisition rate, data acquired during one walking or hand gesture are numerous and can contain redundant data. To overcome the above problem, and long-term dependency issues which are common with RNNs, an LSTM network is better suited for gesture recognition. It will reduce the gradient vanishing problem over time with separate memory cells that carry information over many time steps. Since the pattern that is wanted to be recognized is broad and simple, a not-too-deep network can be used with less complexity. Hence a network with LSTM cells was decided to be used in the gesture recognition problem. The network architecture used for raw accelerometer and gyroscope data is shown in \ref{fig:lstm_code}

\subsection{Gait Analysis}
There are many several methods and mathematical models that can be used to extract and analyze gait parameters. For IMU data, inverse dynamics can be used to extract gait parameters \cite{book1}. The first step in inverse dynamics analysis is to measure the kinematics of the body during movement. This data is used to calculate the position, velocity, and acceleration of each segment of the body, as well as the orientation of each joint. This can be achieved by using a kinematic chain model, which represents the body or limb as a series of interconnected segments, each with its own set of joint angles. The joint angles can be calculated using the inverse kinematic equations for each segment in the chain, based on the desired movement of the end effector, such as the foot or hand.

\begin{figure*}[!t]
\centering
\begin{subfigure}[t]{0.2\textwidth}
    \includegraphics[width=1\textwidth]{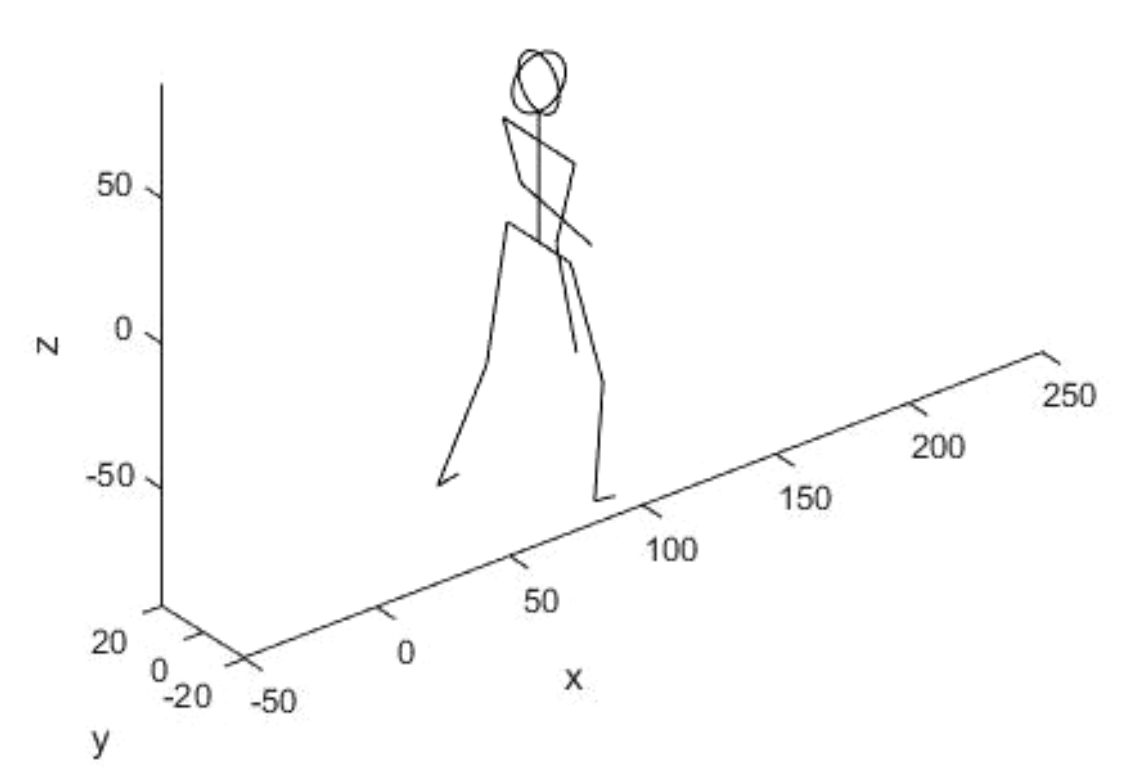}  \caption{A}
        \label{fig:A}
\end{subfigure}
\begin{subfigure}[t]{0.2\textwidth}
    \includegraphics[width=1\textwidth]{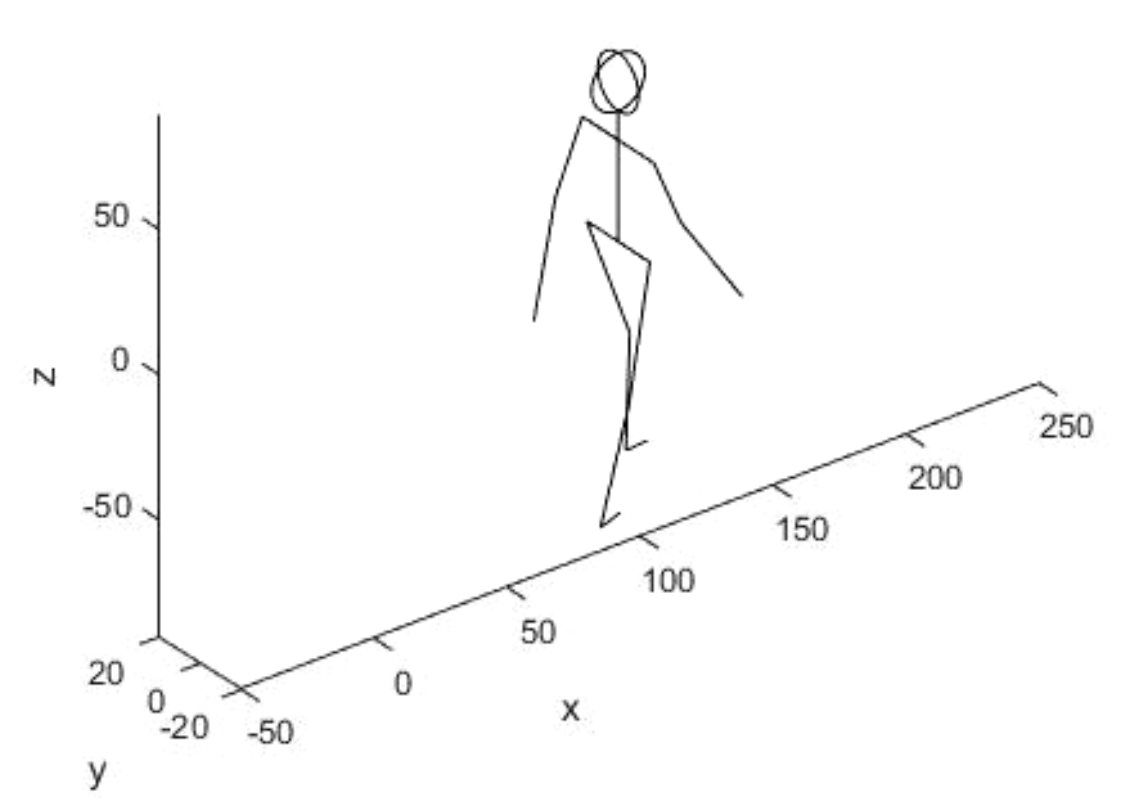}
        \caption{B}
        \label{fig:B}
\end{subfigure}
\begin{subfigure}[t]{0.2\textwidth}
    \includegraphics[width=1\textwidth]{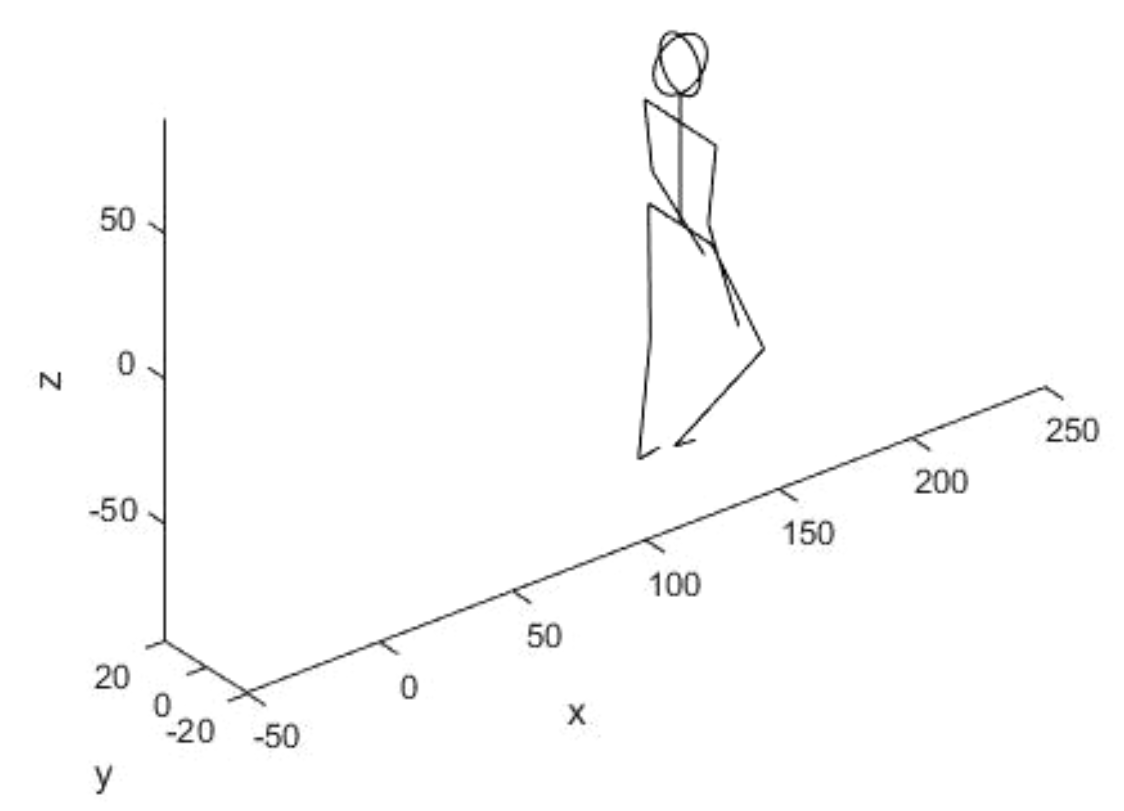}
        \caption{C}
        \label{fig:B}
\end{subfigure}
\caption{Visualization through Gait Parameter extraction}
\label{fig:gait}
\end{figure*}

\begin{table}[!t]
    \centering
    \caption{Performance of the Machine learning techniques for the classification of hand movement}
    \label{tab:results_hand}
    \begin{tabular}{c c c c c| c c c c }
       \multicolumn{5}{c|}{Stages Used}  & \multicolumn{4}{c}{Algorithms} \\
       S1 & S2 & S3 & S4 & S5 & LR & KNN & RF & NB\\
       \hline\hline
       \checkmark & \checkmark & & & \checkmark & 21.28 & 60.64 & \textbf{74.16} & 68.09\\
       \checkmark & \checkmark & & \checkmark & \checkmark & 62.92 & 61.80 & 68.54 & \textbf{75.53}\\
       & & \checkmark & & \checkmark & 71.28 & 75.53 & 84.04 & \textbf{84.05}\\
       & & \checkmark & \checkmark & \checkmark & \textbf{92.14} & 79.78 & 89.89 & 80.90\\
    \end{tabular}
\end{table}

\begin{table}[!t]
    \centering
    \caption{Performance of the Machine learning for the classification of walking patterns}
    \label{tab:results_walking}
    \begin{tabular}{c c c c c| c c c c}
       \multicolumn{5}{c|}{Stages Used}  & \multicolumn{4}{c}{Algorithms} \\
       S1 & S2 & S3 & S4 & S5 & LR & KNN & RF & NB\\
       \hline\hline
        \checkmark & \checkmark & & \checkmark & \checkmark & \textbf{68.18} & 56.06 & 54.55 & 54.85\\
       & & \checkmark & \checkmark & \checkmark & 71.02 & 74.56 & 72.81 & \textbf{76.24}\\
    \end{tabular}
\end{table}

\begin{figure}[!t]
\centering
\begin{subfigure}[t]{0.22\textwidth}
    \includegraphics[width=1\textwidth]{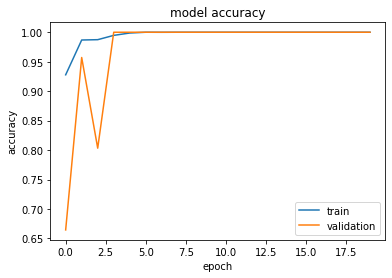} 
        \caption{Accuracy vs Epochs}
        \label{fig:A}
\end{subfigure}
\begin{subfigure}[t]{0.22\textwidth}
    \includegraphics[width=1\textwidth]{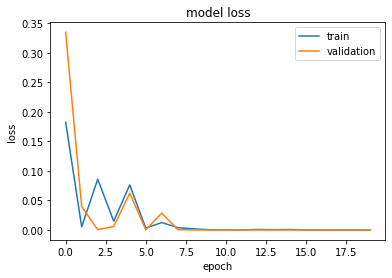}
        \caption{Loss vs Epochs}
        \label{fig:B}
\end{subfigure}
\caption{Hand Movements LSTM Model Accuracy and Losses}
\label{fig:hand_lstm}
\end{figure}

\begin{figure}[!t]
\centering
\begin{subfigure}[t]{0.22\textwidth}
    \includegraphics[width=1\textwidth]{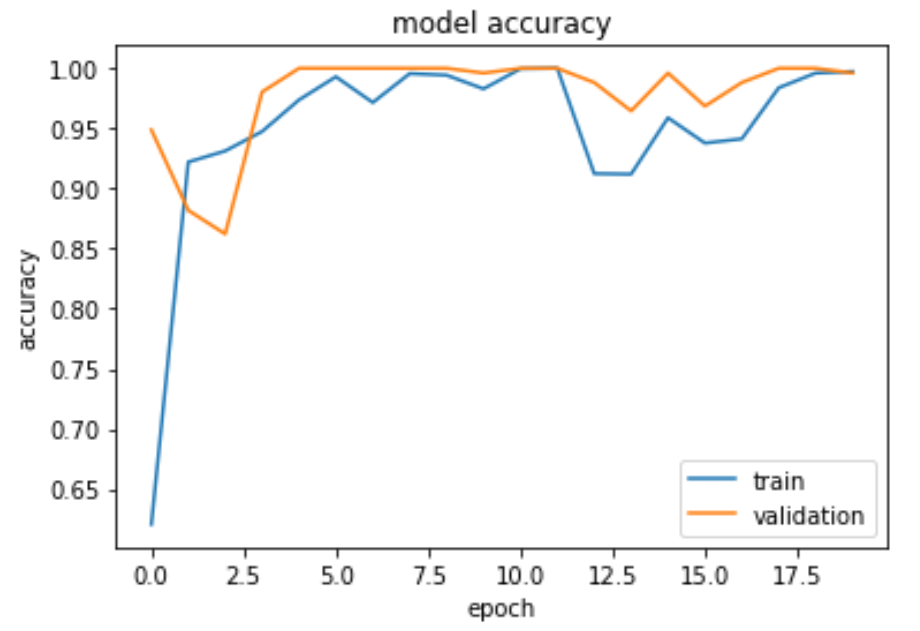}  \caption{Accuracy vs Epochs}
        \label{fig:A}
\end{subfigure}
\begin{subfigure}[t]{0.22\textwidth}
    \includegraphics[width=1\textwidth]{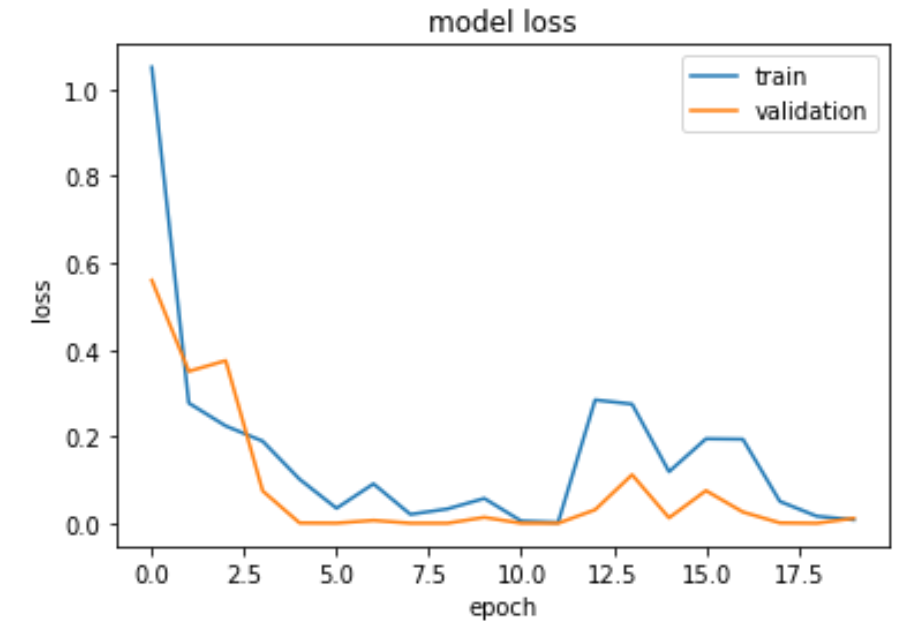}
        \caption{Loss vs Epochs}
        \label{fig:B}
\end{subfigure}
\caption{Walking Patterns LSTM Model Accuracy and Loss}
\label{fig:walk_lstm}
\end{figure}

\section{Results} 
This section provides the experimental results of the design of the wearable device, robust data acquisition method, and data analysis. As far as the outcomes of the design of the wearable contraption are concerned, this work managed to compose an easily wearable, less technical oriented, versatile device that basically captures the particular set of movements. In terms of data acquisition, this work successfully utilizes the existing resources of the IMUs and other electronic devices that are used to implement this device. 

Finally, the analysis of the data captured by the designed device was compared in Table \ref{tab:results_hand} and Table \ref{tab:results_walking}.
From the ML algorithm, LR gave the highest accuracy of  $92.14$\%  for classifying hand movements and NB gave the highest accuracy of $76.24$\% for walking patterns. Both these cases have followed the same set of sensor filtering processes of S3, S4, and S5. This proves that a combination of the aforementioned steps can eliminate the correlation between the data and activating the distance-based feature space has directly contributed to extracting the most important features from the data. With closer inspection, it can be observed that accuracies of LR, KNN, and NB increased with the introduction of distance measure-based dimension reduction, while it has further improved with the introduction of the Kalman filter in the place of signal power-based and clustering measurement-based filtering.  

LSTM deep learning algorithm provided a $100$\% accuracy in terms of classification in both hand movements and walking patterns. Figure \ref{fig:walk_lstm} and Figure \ref{fig:hand_lstm} show the model accuracy and model loss, against the number of epochs.

As an extension of this work, estimating, analyzing, and visualizing the gait parameters will be carried on. Figure \ref{fig:gait} shows the visualization using some of the gait parameters generated from the data acquired by the designed wearable device.
      
In situations where more complex motion estimation tasks and gait estimation tasks need to be analyzed, DL algorithms may not perform perfectly. However, in a similar fashion to the case studies done with ML, using multi-stage data processing combinations might incrementally improve performance in DL algorithms. This can be examined in future research activities.

\section{Conclusion}

The designed device managed to captivate measurements and identify even slight changes in hand movements and walking patterns. In addition, the proposed device and the developed algorithms have the capability of optimizing the data acquisition through IMU sensors and classifying the actions accurately through ML and DL techniques. The 100\% accuracy obtained in the LSTM network has proved the data capturing capability and quality of the data captured from the device. 

In conclusion, the device was simple enough to be operated even by nontechnical personnel while it has the capability to capture subtle movements associated with limb movements. The robust data acquisition systems utilized readily available resources such as IMUs, Arduino nano, and SD modules. The modular and nonrestrictive nature of the wearable device allowed capturing of data ranging from singular limb movements to whole-body movements in uncontrolled environments. Due to this design architecture, it was able to capture the effects of minute movements and subtle changes in the same human motion patterns due to external factors such as the environment and accessories such as shoes.

The proposed active filtering methods in this article have a significant impact on the accuracy of ML algorithms. While for unsmoothed data addition of distance measure-based dimension reduction technique labeled as S4, yield mixed results as shown in Table \ref{tab:results_hand}. When the data is smoothened and represented through the Kalman filter, the introduction of S4 has a significant impact on the performance. Based on the results obtained in the above steps, it can be inferred that the Kalman filter functions as a noise reduction step, while the S4 acts as a temporal dimension reduction step, successfully compressing the classification data for use in ML algorithms.

The LSTM neural network used in this study achieved a flawless accuracy of 100\% in classifying human motion. This validates that the wearable device provided reasonably solid raw data for a lightweight neural network to be operated on to classify subtle motion pattern changes. 

In conclusion, this device which is lightweight, nonobstructive, and can be dedicated for long-term use for a given worker is an effective tool to monitor, and identify activities done by the worker and the impact of environmental change on a worker and thereafter detect subtle changes in gait parameters to detect the possibility of work-related MSD.

\balance
\bibliographystyle{IEEEtran}
\bibliography{bibliography}

\begin{thebibliography}{10}
\providecommand{\url}[1]{#1}
\csname url@samestyle\endcsname
\providecommand{\newblock}{\relax}
\providecommand{\bibinfo}[2]{#2}
\providecommand{\BIBentrySTDinterwordspacing}{\spaceskip=0pt\relax}
\providecommand{\BIBentryALTinterwordstretchfactor}{4}
\providecommand{\BIBentryALTinterwordspacing}{\spaceskip=\fontdimen2\font plus
\BIBentryALTinterwordstretchfactor\fontdimen3\font minus
  \fontdimen4\font\relax}
\providecommand{\BIBforeignlanguage}[2]{{%
\expandafter\ifx\csname l@#1\endcsname\relax
\typeout{** WARNING: IEEEtran.bst: No hyphenation pattern has been}%
\typeout{** loaded for the language `#1'. Using the pattern for}%
\typeout{** the default language instead.}%
\else
\language=\csname l@#1\endcsname
\fi
#2}}
\providecommand{\BIBdecl}{\relax}
\BIBdecl

\bibitem{HMA}
J.~K. Aggarwal and Q.~Cai, ``Human motion analysis: A review,'' \emph{Computer
  vision and image understanding}, vol.~73, no.~3, pp. 428--440, 1999.

\bibitem{HMAmedi}
B.~Najafi, K.~Aminian, A.~Paraschiv-Ionescu, F.~Loew, C.~J. Bula, and
  P.~Robert, ``Ambulatory system for human motion analysis using a kinematic
  sensor: monitoring of daily physical activity in the elderly,'' \emph{IEEE
  Transactions on biomedical Engineering}, vol.~50, no.~6, pp. 711--723, 2003.

\bibitem{hma_sports}
J.~S. Kim, ``Dnn-based human activity recognition by learning initial motion
  data for virtual multi-sports,'' vol.~23, pp. 373--375, 01 2021.

\bibitem{hma_human_machine}
A.~M. Sabanchiev, R.~S. Nakhushev, A.~A. Chetvertakov, and M.~I. Aliev,
  ``Analysis of the human-machine interface for controlling a robotic
  manipulator using gestures,'' in \emph{2020 International Conference Quality
  Management, Transport and Information Security, Information Technologies
  (IT\&QM\&IS)}, 2020, pp. 341--344.

\bibitem{hma_surveillance}
W.~Lao, J.~Han, and P.~H. De~With, ``Automatic video-based human motion
  analyzer for consumer surveillance system,'' \emph{IEEE Transactions on
  Consumer Electronics}, vol.~55, no.~2, pp. 591--598, 2009.

\bibitem{hma_physio}
W.~Wei, C.~McElroy, and S.~Dey, ``Towards on-demand virtual physical therapist:
  Machine learning-based patient action understanding, assessment and task
  recommendation,'' \emph{IEEE Transactions on Neural Systems and
  Rehabilitation Engineering}, vol.~27, no.~9, pp. 1824--1835, 2019.

\bibitem{inproceedings}
M.~Vasconcelos and J.~Tavares, ``Human motion analysis: Methodologies and
  applications,'' 02 2008.

\bibitem{IMUVsOptical}
Y.-J. Lu, C.-J. Chang, C.-W. Chang, and S.-W. Yang, ``Accuracy comparisons in
  imu sensor and motion analysis software,'' in \emph{Proceedings of the 2018
  2nd International Conference on Mechatronics Systems and Control
  Engineering}, 2018, pp. 13--16.

\bibitem{IMUSinVsMul}
L.~Gao, A.~Bourke, and J.~Nelson, ``Evaluation of accelerometer based
  multi-sensor versus single-sensor activity recognition systems,''
  \emph{Medical engineering \& physics}, vol.~36, no.~6, pp. 779--785, 2014.

\bibitem{military}
M.~P. Mavor, G.~B. Ross, A.~L. Clouthier, T.~Karakolis, and R.~B. Graham,
  ``Validation of an imu suit for military-based tasks,'' \emph{Sensors},
  vol.~20, no.~15, 2020.

\bibitem{Sports}
J.~Lee, H.~Espinosa, and D.~James, ``The inertial sensor: A base platform for
  wider adoption in sports science applications,'' \emph{Journal of Fitness
  Research}, vol.~4, 01 2015.

\bibitem{safety}
Y.~Kim, H.~Jung, B.~Koo, J.~Kim, T.~Kim, and Y.~Nam, ``Detection of pre-impact
  falls from heights using an inertial measurement unit sensor,''
  \emph{Sensors}, vol.~20, no.~18, 2020.

\bibitem{MSDcashew}
N.~Girish, K.~Ramachandra, M.~Arun~G, and K.~Asha, ``Prevalence of
  musculoskeletal disorders among cashew factory workers,'' \emph{Archives of
  environmental \& occupational health}, vol.~67, no.~1, pp. 37--42, 2012.

\bibitem{MSDchina}
W.~Yu, T.~Ignatius, Z.~Li, X.~Wang, T.~Sun, H.~Lin, S.~Wan, H.~Qiu, and S.~Xie,
  ``Work-related injuries and musculoskeletal disorders among factory workers
  in a major city of china,'' \emph{Accident Analysis \& Prevention}, vol.~48,
  pp. 457--463, 2012.

\bibitem{MSDiran}
A.~Choobineh, S.~H. Tabatabaei, A.~Mokhtarzadeh, and M.~Salehi,
  ``Musculoskeletal problems among workers of an iranian rubber factory,''
  \emph{Journal of occupational health}, vol.~49, no.~5, pp. 418--423, 2007.

\bibitem{MSDiranSugar}
A.~Choobineh, S.~H. Tabatabaee, and M.~Behzadi, ``Musculoskeletal problems
  among workers of an iranian sugar-producing factory,'' \emph{International
  journal of occupational safety and ergonomics}, vol.~15, no.~4, pp. 419--424,
  2009.

\bibitem{MSDsrilanka}
S.~R. Lombardo, P.~Vijitha~de Silva, H.~J. Lipscomb, and T.~{\O}stbye,
  ``Musculoskeletal symptoms among female garment factory workers in sri
  lanka,'' \emph{International journal of occupational and environmental
  health}, vol.~18, no.~3, pp. 210--219, 2012.

\bibitem{MSDnurse}
B.~Saberipour, S.~Ghanbari, K.~Zarea, M.~Gheibizadeh, and M.~Zahedian,
  ``Investigating prevalence of musculoskeletal disorders among iranian nurses:
  A systematic review and meta-analysis,'' \emph{Clinical Epidemiology and
  Global Health}, vol.~7, no.~3, pp. 513--518, 2019.

\bibitem{MSDvsSafetyE}
J.~O. Crawford, D.~Berkovic, J.~Erwin, S.~M. Copsey, A.~Davis, E.~Giagloglou,
  A.~Yazdani, J.~Hartvigsen, R.~Graveling, and A.~Woolf, ``Musculoskeletal
  health in the workplace,'' \emph{Best practice \& research clinical
  rheumatology}, vol.~34, no.~5, p. 101558, 2020.

\bibitem{IMUbest}
A.~I. Cuesta-Vargas, A.~Gal{\'a}n-Mercant, and J.~M. Williams, ``The use of
  inertial sensors system for human motion analysis,'' \emph{Physical Therapy
  Reviews}, vol.~15, no.~6, pp. 462--473, 2010.

\bibitem{IEEEtrans1}
C.~Yi, S.~Rho, B.~Wei, C.~Yang, Z.~Ding, Z.~Chen, and F.~Jiang, ``Detecting and
  correcting imu movements during joint angle estimation,'' \emph{IEEE
  Transactions on Instrumentation and Measurement}, vol.~71, pp. 1--1, 01 2022.

\bibitem{loop1}
S.~Mohammed~Ali, B.~George, and L.~Vanajakshi, ``An efficient multiple-loop
  sensor configuration applicable for undisciplined traffic,''
  \emph{Intelligent Transportation Systems, IEEE Transactions on}, vol.~14, pp.
  1151--1161, 09 2013.

\bibitem{loop2}
T.~Brunner, S.~Changey, J.-P. Lauffenburger, and M.~Basset, ``Multiple mems-imu
  localization: Architecture comparison and performance assessment,''
  \emph{22nd Saint Petersburg International Conference on Integrated Navigation
  Systems, ICINS 2015 - Proceedings}, pp. 123--126, 01 2015.

\bibitem{heel1}
J.~Maduabuchi, N.~Joseph, A.~Egwuonwu, A.~O. Ezeukwu, and C.~Nwafulume,
  ``Effects of different heel heights on selected gait parameters of young
  undergraduate females,'' \emph{Page Header Logo Journal of Paramedical
  Sciences}, vol.~3, pp. 2008--4978, 01 2012.

\bibitem{heel2}
M.~L.~J. C.~Basilan, {https://orcid.org/0000-0003-3105-2252}, M.~Padilla, and
  {https://orchid.org/0000-0001-5025-12872, maleticiajose.basilan@deped.gov.ph,
  maycee.padilla@deped.gov.ph, Department of Education- SDO Batangas Province,
  Batangas, Philippines}, ``Assessment of teaching english language skills:
  Input to digitized activities for campus journalism advisers,''
  \emph{International Multidisciplinary Research Journal}, vol.~4, no.~4, Jan.
  2023.

\bibitem{intra_inter1}
S.~Wu and T.~Chow, ``Clustering of the self-organizing map using a clustering
  validity index based on inter-cluster and intra-cluster density,''
  \emph{Pattern Recognition}, vol.~37, pp. 175--188, 2004.

\bibitem{kalman_for_attitude}
E.~J. Lefferts, F.~L. Markley, and M.~D. Shuster, ``Kalman filtering for
  spacecraft attitude estimation,'' \emph{Journal of Guidance, Control, and
  Dynamics}, vol.~5, no.~5, pp. 417--429, 1982.

\bibitem{book1}
J.~Garza-Ulloa, \emph{Book: Applied Biomechatronics Using Mathematical Model},
  06 2018.

\end{thebibliography}

\end{document}